\documentclass[journal]{IEEEtran}
\usepackage{array}
\usepackage{url}
\usepackage{cite,graphicx,multirow,multicol,setspace,amsmath,wrapfig}%,subcaption}
\usepackage{tabulary,amssymb,amsfonts,adjustbox,booktabs,qtree,tikz-qtree,synttree}
\usepackage[flushleft]{threeparttable}
\usepackage{makecell,eqnarray}
\usepackage{cuted}
\usepackage[linesnumbered,ruled]{algorithm2e}
\usepackage{algorithmic,textcomp,xcolor}
\usepackage{pifont}
\newcommand{\cmark}{\ding{51}}%
\usepackage{commath}
\usepackage[switch]{lineno}

\graphicspath{ {Images/} }

\makeatletter
\let\MYcaption\@makecaption
\makeatother

\usepackage[font=footnotesize]{subcaption}

\makeatletter
\let\@makecaption\MYcaption
\makeatother

\def\BibTeX{{\rm B\kern-.05em{\sc i\kern-.025em b}\kern-.08em
    T\kern-.1667em\lower.7ex\hbox{E}\kern-.125emX}}
\usepackage{amsthm}
\theoremstyle{definition}
\newtheorem{definition}{Definition}%[section]
\makeatletter
\newcommand{\algorithmfootnote}[2][\footnotesize]{%
  \let\old@algocf@finish\@algocf@finish% Store algorithm finish macro
  \def\@algocf@finish{\old@algocf@finish% Update finish macro to insert "footnote"
    \leavevmode\rlap{\begin{minipage}{\linewidth}
    #1#2
    \end{minipage}}%
  }%
}

\begin{document}
% \linenumbers
\title{Multi-Objective Evolutionary Approach to Grey-Box Identification of Buck Converter}

% \title{Multi-objective Evolutionary Approach to Grey-Box Identification: A Case Study of Buck Converter}
%
\author{Faizal~Hafiz,
        Akshya~Swain,
        Eduardo~M.~A.~M.~Mendes and~Luis~A.~Aguirre~\IEEEmembership{}% <-this % stops a space
\thanks{F. Hafiz and A. Swain are with the Department of Electrical \& Computer Engineering, The University of Auckland, Auckland, New Zealand. E-mail: faizalhafiz@ieee.org, a.swain@auckland.ac.nz.}% <-this % stops a space

\thanks{E. Mendes and L. Aguirre are with the Department of Electronics Engineering, Federal University of Minas Gerais, Belo Horizonte, Brazil. Email : emmendes@cpdee.ufmg.br and aguirre@cpdee.ufmg.br. They acknowledge financial support from CNPq, CAPES and FAPEMIG (Brazil).}% <-this % stops a space
\thanks{This article has been accepted for publication in IEEE Transaction on Circuits and Systems-I: Regular Papers. The published version is available at DOI: 10.1109/TCSI.2020.2970759}}

% \markboth{IEEE TRANSACTIONS ON CIRCUITS AND SYSTEMS-I: REGULAR PAPERS}%
% {Hafiz \MakeLowercase{\textit{et al.}}: Multi-objective Evolutionary Approach to Grey-Box Identification of Buck Converter}


\maketitle

\begin{abstract}

The present study proposes a simple \textit{grey-box} identification approach to model a real DC-DC buck converter operating in continuous conduction mode. The problem associated with the information void in the observed dynamical data, which is often obtained over a relatively narrow input range, is alleviated by exploiting the known static behavior of buck converter as \textit{a priori} knowledge. A simple method is developed based on the concept of term clusters to determine the static response of the candidate models. The error in the static behavior is then directly embedded into the multi-objective framework for structure selection. In essence, the proposed approach casts grey-box identification problem into a multi-objective framework to balance \textit{bias-variance} dilemma of model building while explicitly integrating \textit{a priori} knowledge into the \textit{structure selection} process. The results of the investigation, considering the case of practical buck converter, demonstrate that it is possible to identify parsimonious models which can capture both the dynamic and static behavior of the system over a wide input range.

\end{abstract}


\ifCLASSOPTIONpeerreview
\begin{center} \bfseries EDICS Category: 3-BBND \end{center}
\fi
\IEEEpeerreviewmaketitle

\section{Introduction}

% ADD 2-3 lines: DC/DC modeling - small signal,circuit level vs. system identification....\\

\IEEEPARstart{M}{odeling} is the first step for control, condition-monitoring and fault diagnosis of power electronic converters. Over the past few years, this field has attracted significant research attention, which range from circuit topology and linear analysis based modeling approaches to data-driven modeling such as system identification and neural networks~\cite{Al-Greer:Armstrong:2019}. Among these, data-driven modeling is particularly well suited to handle inherent non-linearities of converters and can successfully account for uncertainties associated with stray parameter changes and aging effects. This study, therefore, follows a system identification based approach to model DC-DC buck converter operating in continuous conduction mode.

System identification deals with the development of mathematical descriptors of system dynamics from the observed dynamical data~\cite{Schoukens:Ljung:2019,Billings:2013,Aguirre:2019}. To this end, it is essential that the system under investigation is persistently excited over wide operating conditions so that the system dynamics are captured in the observed data, and subsequently encoded into the identified model. However, in practice, it is often difficult to drive the system over a wide input range. In such a scenario, the model identified using only observed dynamical data may not generalize well as the observed data contains information over a relatively small range of system dynamics. In this study, we consider a practical case study of modeling buck converter, which falls under this category. In particular, the buck converter considered here is excited over a relatively narrow input range. Our previous investigations~\cite{Aguirre:Donoso:2000,Correa:Aguirre:2002} on this case study show that while the models identified following \textit{black-box} identification (\textit{i.e.}, using only dynamical data) can capture the converter dynamics, they cannot preserve the static non-linearity of the converter beyond local input range. In this study, we propose possible remedies based on the philosophy of grey-box identification to aid the identification process in such scenarios.

When the observed dynamical data contain only limited information about the system behavior, the identification process can be augmented by including additional auxiliary information about the system under investigation which could either be obtained by \textit{first principle} or steady state data, \textit{e.g.}, static function, number and location of fixed points~\cite{Johansen:1996,Erivelton:Takahashi:2003,Nepomuceno:Takahashi:2007,Aguirre:Barroso:2004,Barbosa:Aguirre:2011,Aguirre:Donoso:2000,Correa:Aguirre:2002,Aguirre:2019,Martins:Nepomuceno:2013}. This auxiliary information, often referred to as \textit{a priori knowledge}, can provide vital information about system behavior and can aid the identification process. Given that only finite data points are available for the identification, any \textit{a priori} knowledge about the system under consideration is a welcome feature. The focus of this study is, therefore, the \textit{grey-box} identification approach, which explicitly utilizes such \textit{a priori} knowledge.

The major challenge of grey-box identification is to develop a suitable framework that can articulate and embed \textit{a priori} system knowledge into the identification process. To this end, such \textit{a priori} knowledge can be integrated into either of the following steps of the model building process: 1) \textit{Structure Selection} and 2) \textit{Parameter Estimation}. Given that most of the system representations such as Volterra and Nonlinear Auto-regressive with eXogenous inputs (NARX) are \textit{linear-in-parameter}, the parameters of such models can be estimated following least-squares based algorithms. In contrast, structure selection, which involves the identification of significant \textit{terms/basis functions}, is a much more complex issue, and it is one of the fundamental problems of system identification. It is easy to follow that, the `\textit{quality}' of the identified \textit{grey}-models can significantly be improved if \textit{a priori} knowledge is directly integrated into the fundamental step of structure selection. However, to the best of our knowledge, this issue is yet to be explored in grey-box identification. This has been the main motivation for this study.

In most of the existing grey-box identification approaches, it is assumed that the structure of the system under consideration is known, and \textit{a priori} knowledge is embedded into the parameter estimation. For instance, \textit{a priori} information about static gain and fixed point is utilized to constrain the estimated parameters in~\cite{Correa:Aguirre:2002,Erivelton:Takahashi:2003,Nepomuceno:Takahashi:2007,Aguirre:Barroso:2004}. In~\cite{Barbosa:Aguirre:2011}, the parameter estimation is formulated as a bi-objective problem to incorporate known steady-state behavior of the system. A detailed treatment of such grey-box identification approaches can be found  in~\cite{Aguirre:2019}. Further, a few notable exceptions to parameter estimation based approaches can be found in~\cite{Aguirre:Donoso:2000,Martins:Nepomuceno:2013}. In~\cite{Aguirre:Donoso:2000}, the pool of viable system terms is restricted beforehand, based on a known static gain of the system. This approach, however, involves a trade-off in the dynamic prediction capabilities. In~\cite{Martins:Nepomuceno:2013}, Martins \textit{et al.} extended the well-known Error-Reduction-Ratio (ERR) metric~\cite{Billings:2013} to incorporate the \textit{a priori} knowledge into the structure selection. In particular, the ERR metric is determined from both the dynamical observations as well as steady-state information. Finally, the \textit{weighted}-sum of these metrics is utilized to select the system structure. While this approach is attractive, it is often cumbersome to determine the appropriate weights \textit{a priori}. Consequently, the entire identification procedure has to be repeated with a distinct combination of weights. Further, this approach is dependent on an auxiliary routine to determine the order of the model (number of terms), which is crucial to balance the bias-variance dilemma in system identification~\cite{Billings:2013,Hafiz:Swain:2DUPSO:2019,Hafiz:Swain:MOEA:2020}.

To the best of our knowledge, the explicit use of \textit{a priori} knowledge for the benefit of structure selection is still an open issue in the grey-box identification. To bridge this gap, a simple and effective approach to embed  \textit{a priori} system knowledge directly into the structure selection process is proposed in this study. The \textit{a priori} knowledge is quantified and explicitly formulated as one of the search objectives of the multi-objective structure selection procedure. This study essentially proposes a Multi-criteria Decision Making (MCDM) framework for grey-box identification, which is the combination of a Multi-objective Evolutionary Algorithm (MOEA) and \textit{a posteriori} preference articulation technique. The following are the key contributions of this investigation:

\begin{itemize}

    \item A simple approach is developed based on the concept of `\textit{term-clusters}'~\cite{Aguirre:Billings:1995b} to quantify \textit{a priori} knowledge. In essence, the proposed approach can determine and compare the \textit{static} response of the \textit{candidate} models with \textit{a priori} knowledge, which can easily be exploited by a structure selection algorithm. Given that the static information can be obtained with relative ease (\textit{e.g., through steady-state measurements}), this approach can be used to extrapolate the information contained in the observed dynamical data.

    \item The \textit{bias-variance} dilemma is one of the fundamental issues of structure selection; \textit{under-fitted} models fail to capture the system dynamics, whereas \textit{over-fitted} models may introduce undesired dynamics which are not present in the original system~\cite{Aguirre:Billings:1995}. This study convincingly demonstrates that it is possible to balance the bias-variance dilemma while embedding the \textit{a priori} knowledge into the structure selection. It is shown that this can be achieved by classical MOEAs such as NSGA-II and SPEA-II~\cite{Deb:Pratap:2002,Zitzler:Laumanns:2001}. 
 
    % \item It is known that discrete-time model of a continuous-time system is not unique~\cite{}. It is demonstrated that the proposed approach can identify multiple valid discrete-time description for such systems. %
    
    % \item While there may exist multiple valid discrete-time models, in practice, the Decision-Maker (DM) is tasked with the selection of a few models. Hence, a simple model ranking procedure is proposed for the final selection of a few models from the identified non-dominated models.

\end{itemize}

The efficacy of the proposed approach is demonstrated by a practical case study on a DC-DC buck converter operating in the continuous conduction mode, which was reported earlier by the authors in~\cite{Aguirre:Donoso:2000,Correa:Aguirre:2002}. The challenge here is to identify globally valid models as the converter excited with a relatively narrow range of input. Consequently, the observed dynamic data do not contain enough information to mimic the converter behavior beyond a local input range. This information void can be supplemented by the known static curve of buck converter, \textit{i.e.}, the first principle relationship between Pulse Width Modulation (PWM) DC voltage and the converter output voltage. Based on this notion, our earlier attempts to exploit this \textit{a priori} knowledge in grey-box identification were reported in~\cite{Aguirre:Donoso:2000,Correa:Aguirre:2002}. While these earlier approaches could identify globally valid models, these often involve a trade-off in the dynamic prediction capabilities. In contrast, this study proposes the use of \textit{a priori} knowledge at the fundamental level of structure selection, and it is essentially a further step in grey-box identification. This is convincingly demonstrated by a detailed comparative evaluation on the same case study.

The rest of the article is organized as follows: The experimental setup to gather identification data from the buck converter is described in Section~\ref{s:buck}. The polynomial NARX model, term clusters and the structure selection problem are discussed briefly in Section~\ref{s:prelim}. The proposed multi-objective structure selection approach is discussed in detail in Section~\ref{s:propsedapproach}. The results are discussed at length in Section~\ref{s:res}, followed by the conclusions in Section~\ref{s:con}.

% In what follows, we discuss a practical case study on modeling of DC-DC converter 

% This study, in particular, focuses on modeling of DC-DC buck converter operating in continuous conduction mode.

% CASE STUDY ON BUCK CONVERTER - LIMITATIONS - IDENTIFICATION DATA OVER SMALL RANGE---
% In the example, we should emphasize that the data that contains the dynamic in-formation cover a small range of the system dynamics.  Without the static data, a model is very likely not to generalize well.  Since the static information is usually easy to obtain, our approach can be used to find models that extrapolates the dynamics in the dynamic data.\\
% TRANSITION TO GREY BOX\\To this end, the given system can be represented by Wiener, Hammerstein, Neural Networks, Nonlinear Auto-Regressive with eXogeneous inputs (NARX) models and others. This study focuses on NARX model due to its convenient linear-in-parameter form and associated frequency domain analysis tools~\cite{}.  

%-------------------------------------------------------------
\section{Modelling of DC-DC Converter Dynamics}
\label{s:buck}

The objective of this study is to find a nonlinear model which successfully captures the dynamic behavior of the buck converter. The identification data for this purpose is gathered from the experimental setup described in Section~\ref{s:data}. Further, the static behavior of the buck converter is known. The use of this \textit{a priori} information and the modelling objectives are discussed in Section~\ref{s:obj}.
%%---- Data --------------------------------------------------
\begin{figure}[!b]
\centering
\small
  \includegraphics[width=0.43\textwidth]{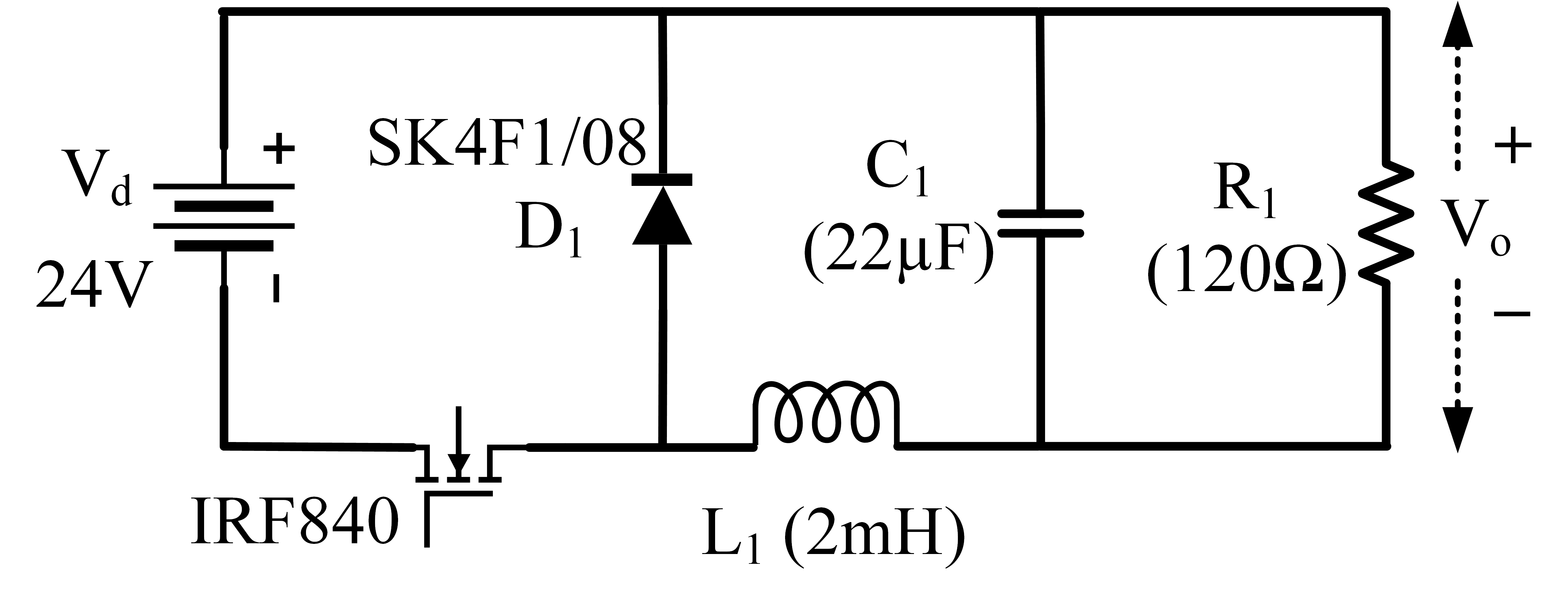}
\caption{The buck converter considered in this study. The converter is driven by MOSFET IRF840. The PWM switching is controlled by LM3524 at $33kHz$.}
\label{f:buck}
\end{figure}
%----------------------------------------------------------------------
%%---- Data --------------------------------------------------
\begin{figure}[!t]
\centering
\small
\begin{subfigure}{.31\textwidth}
  \centering
  \includegraphics[width=\textwidth]{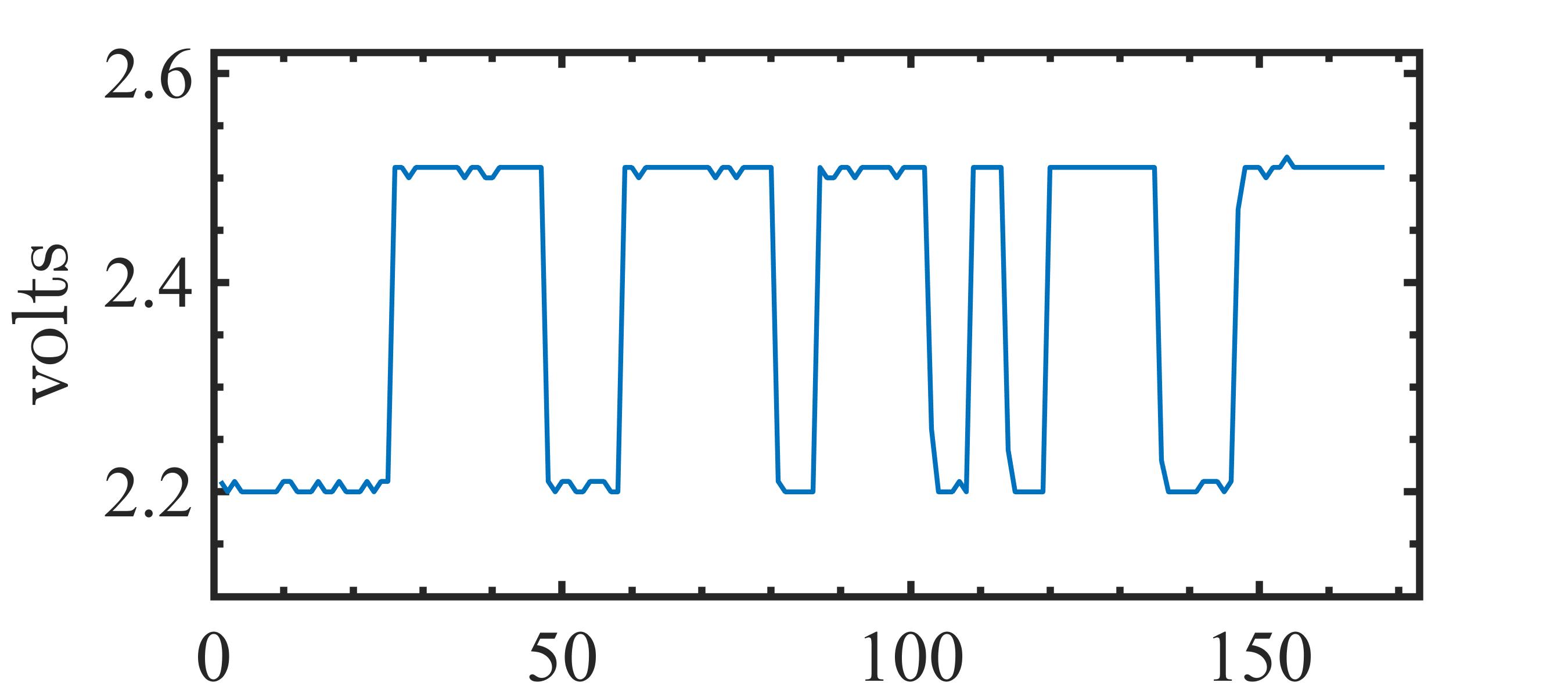}
  \caption{$u(k)$}
  \label{f:umeas}
\end{subfigure}
\hfill
\begin{subfigure}{0.31\textwidth}
  \centering
  \includegraphics[width=\textwidth]{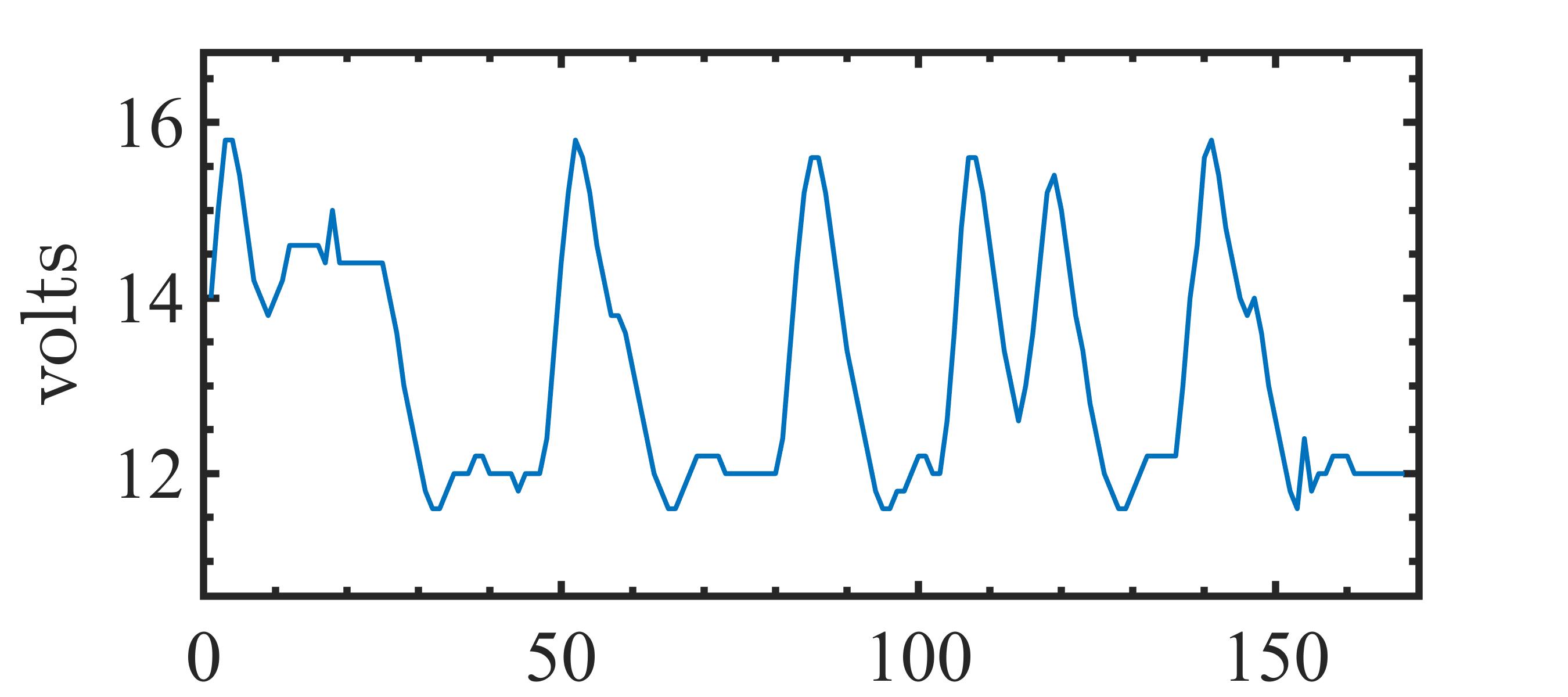}
  \caption{$y(k)$}
  \label{f:ymeas}
\end{subfigure}
\caption{The identification data considered in this study. (a) Model Input ($u$): PWM DC voltage (c) Model Output ($y$): Buck output voltage}
\label{f:iddata}
\end{figure}
%----------------------------------------------------------------------

%-------------------------------------------------------------
\subsection{Data Acquisition}
\label{s:data}

In this study, a buck converter operating in the \textit{continuous} conduction mode is considered. For this purpose, the buck converter is implemented as shown in Fig.~\ref{f:buck}. The input voltage, `$V_d$', is regulated at $24 V$ throughout the experiment. The output voltage, `$V_o$', is controlled by the Pulse Width Modulation (PWM) switching of the MOSFET (IRF840). In the PWM, a signal level dc voltage, `$V_{control}$', is compared to a triangular waveform to adjust the duty ratio, $D=\frac{T_{ON}}{T_s}$, as per the prevailing requirements. This is accomplished by the PWM controller (LM3524, \textit{not shown here}) at the rate of $\frac{1}{T_s}=33 kHz$ to ensure the operation in the \textit{continuous} conduction mode, \textit{i.e.}, the current through the inductor `$L_1$' (Fig.~\ref{f:buck}) is never zero.

The main objective of this study is to capture the nonlinear dynamics of the output voltage $V_o$, which is dependent on the duty ratio `$D$' and the consequent energy exchange among $L_1$, $C_1$ and $R_1$ (see Fig.~\ref{f:buck}). For this purpose, a model is identified with the signal level PWM dc voltage, `$V_{control}$', as the \emph{input} (\textit{hereafter denoted by} `$u$') and the voltage $V_o$ as the \emph{output} (\textit{hereafter denoted by} `$y$').

For identification, it is crucial to ensure that the converter is persistently excited so that the essential information about converter dynamics can be gathered. To this end, a Pseudo Random Binary Sequences (PRBS) signal is used as the input, $u$, which drives the converter in the range of $2.2V \leq u(k) \leq 2.5 V$. The consequent changes in the output are captured by a digital oscilloscope at the sampling frequency of $1MHz$. The identification data (shown in Fig.~\ref{f:iddata}) is obtained by decimating the input-output data by a factor of $12$ to avoid the oversampling issues. Further details about the experimental setup and the data acquisition can be found in~\cite{Aguirre:Donoso:2000}.
%-------------------------------------------------------------
\subsection{Modelling Objectives}
\label{s:obj}

The main objective of the identified model is to capture the dynamic behavior of the output voltage. Further, the steady-state relationship between the input and output converter voltages are usually known \emph{a priori}. It is therefore essential to induce such static behavior in the identified models. For example, the steady state voltage relation for the buck converter considered in this study is given by,%~\cite{Aguirre:Donoso:2000}
%-------------------------------------------------------------
\begin{align}
    \label{eq:static}
    \overline{y}=\frac{4V_d}{3} - \frac{V_d}{3} \overline{u}
\end{align}
%-------------------------------------------------------------
where, $\overline{u}$ and $\overline{y}$ respectively denote the steady state values of the input and output. 
It is clear that in addition to a good prediction capability, the identified model must have a steady state relation of the form $\overline{y}=f(\overline{u})$ in order to mimic the static behavior of the buck converter given by~(\ref{eq:static}). This \textit{a priori} information is crucial to the identification process, as will be discussed in Section~\ref{s:useofprior}. 

A \emph{black-box} identification approach is not adequate to achieve the modeling objectives of this study because such an identification approach relies only on the information extracted from the dynamical dataset and the \textit{a priori} information about static behavior is not incorporated. Given that the input drives the system over a relatively narrow range, \textit{i.e.}, $u(k) \in [2.2V, 2.5 V]$, the static behavior of the back-box models is valid only in this \textit{local} input range~\cite{Aguirre:Donoso:2000}. 

Hence, in this study, a \textit{grey-box} identification approach is followed which integrates the \textit{a priori} information about the static behavior of the buck converter (\textit{i.e.}, $\overline{y}=f(\overline{u})$) into the identification process to obtain \textit{globally} valid models, which will be discussed in the following subsections. 

%-------------------------------------------------------------
\section{Preliminaries}
\label{s:prelim}

The first step of the identification is to select system representation amongst many representations, \textit{e.g.}, Volterra, Wiener, Polynomial/Rational Nonlinear Auto-Regressive with eXogenous inputs (NARX), Neural Network and others. This study focuses on the polynomial NARX representation~\cite{Billings:2013}. The rationale behind is two-fold:
\begin{itemize}
    \item The concept of \textit{term-clusters} was originally developed in the context of NARX models~\cite{Aguirre:Billings:1995b}. This concept forms the basis of the proposed grey-box identification approach, as will be discussed in detail in  Section~\ref{s:useofprior} and~\ref{s:MOSS}.
    
    \item The part of this study focuses on comparative evaluation with the earlier investigations in~\cite{Aguirre:Donoso:2000,Correa:Aguirre:2002}, which were also focused on the polynomial NARX models.
\end{itemize}
In the following the polynomial NARX model is briefly discussed in Section~\ref{s:NARX}. Further, the concept of term cluster is essential to derive the static models form the NARX representation, which is discussed briefly in Section~\ref{s:termcluster}.

%-------------------------------------------------------------
\subsection{The Polynomial NARX Model}
\label{s:NARX}

The NARX model represents a non-linear system as a function of recursive lagged input and output terms as follows:
%---------------------------------------------------------
\begin{align*}
y(k) & = F^{n_l} \ \{ \ y(k-1),\ldots,y(k-n_y),\dots \\
     & \; \; \; \ \ \  u(k-1),\ldots, u(k-n_u) \ \}+e(k)
\end{align*}
%---------------------------------------------------------
where $y(k)$ and $u(k)$ respectively represent the output and input at time intervals $k$, $n_y$ and $n_u$ are corresponding lags and $F^{n_l}\{ \cdotp \}$ is some nonlinear function of degree $n_l$. 

The \textit{total number of possible terms} or \textit{model size} ($n$) of the NARX model is given by,

%%---------------------------------------------------------
\begin{align}
\label{eq:Nt}
n & = n_0 + \sum_{i=1}^{n_l} \frac{n_{i-1}(n_y+n_u+i-1)}{i}, \quad n_0=1
\end{align}
%%---------------------------------------------------------
This model is essentially linear-in-parameters and can be expressed as:
%---------------------------------------------------------
\begin{align}
\label{eq:NARXmodel}
    y(k) & = \theta_1 + \sum_{i=2}^{n} \theta_i x_i(k) + e(k)\\
    \text{where, } x_i(k)  & = \prod_{j=1}^{p_y}y(k-n_{y_j})\prod_{k=1}^{q_u}u(k-n_{u_k}) \nonumber
\end{align}
%%---------------------------------------------------------
$p_y,q_u \geq 0$; $1\leq p_y+q_u \leq n_l$; $1 \leq n_{y_j}\leq n_y$;$1 \leq n_{u_k}\leq n_u$; $n_l$ is the degree of polynomial expansion; $k = 1,2,\dots \mathcal{N}$ and `$\mathcal{N}$' denotes the total number of data points.

%-------------------------------------------------------------
\subsection{Term Clusters}
\label{s:termcluster}

The NARX model in (\ref{eq:NARXmodel}) can be represented as summation of terms of with $m^{th}$ order nonlinearity ($1\leq m \leq n_l$) as follows~\cite{Aguirre:Billings:1995b}:
%-------------------------------------------------------------
\begin{align}
\label{eq:cluster1}
    y(k) = & \sum \limits_{m=0}^{n_l}\sum\limits_{p=0}^{m}\sum \limits_{n_1, n_m}^{n_y, n_u} c_{p,m-p} (n_1,\dots, n_m)\nonumber\\
    & \prod\limits_{i=1}^{p} y(k-n_i) \prod \limits_{i=p+1}^{m} u(k-n_i)
\end{align}
%-------------------------------------------------------------
% %-------------------------------------------------------------
% \begin{align*}
%     \text{where, \ }\sum \limits_{n_1,n_m}^{n_y,n_u} = \sum\limits_{n_1=1}^{n_y} \dots \sum \limits_{n_m=1}^{n_u}
% \end{align*}%where, $\sum \limits_{n_1,n_m}^{n_y,n_u} = \sum\limits_{n_1=1}^{n_y} \dots \sum \limits_{n_m=1}^{n_u}$
% %-------------------------------------------------------------
where, $\sum \limits_{n_1,n_m}^{n_y,n_u} = \sum\limits_{n_1=1}^{n_y} \dots \sum \limits_{n_m=1}^{n_u}$ and the upper limit is respectively $n_y$ and $n_u$ for factors $y(k-n_i)$ and $u(k-n_i)$. 

If the model is excited by a constant input and it is asymptotically stable, then the following holds in the steady state,
%-------------------------------------------------------------
\begin{align*}
\overline{y} & =y(k-1)=y(k-2)=\dots=y(k-n_y)\\
\overline{u} & =u(k-1)=u(k-2)=\dots=u(k-n_u)
\end{align*}
%-------------------------------------------------------------
For such condition, (\ref{eq:cluster1}) can further be simplified as follows:
%-------------------------------------------------------------
\begin{align}
\label{eq:cluster2}
    \overline{y} = & \sum \limits_{m=0}^{n_l}\sum\limits_{p=0}^{n_l-m}\sum \limits_{n_1, n_m}^{n_y, n_u} c_{p,m} (n_1,\dots, n_m) \ \overline{y}^p \ \overline{u}^{m}
\end{align}
%-------------------------------------------------------------
%-------------------------------------------------------------
\theoremstyle{definition}
\begin{definition}{\textbf{\textit{Cluster Coefficients}}~\cite{Aguirre:Billings:1995b}:}
The constants $\sum \limits_{n_1, n_m}^{n_y, n_u} c_{p,m} (n_1,\dots, n_m)$ in (\ref{eq:cluster2}) are the coefficients of the \textit{term clusters} $\Omega_{y^{p} u^{m-p}}$, which contain terms of the form $y^{p}(k-i)u^{m}(k-j)$ for $m+p<n_l$. Such coefficients are called \textit{cluster coefficients} and are denoted by $\Sigma_{y^p u^m}$.
\end{definition}
%-------------------------------------------------------------
Following these definitions, the NARX model in the steady state is given by,
%-------------------------------------------------------------
\begin{align}
\label{eq:cluster3}
     \overline{y} = & \Sigma_0 + \Sigma_y \overline{y} + \Sigma_u \overline{u} + \sum \limits_{m=1}^{n_l-1} \sum \limits_{p=1}^{n_l-m} \Sigma_{y^p u^m} \overline{y}^p \ \overline{u}^m \nonumber\\
     & + \sum \limits_{p=2}^{n_l} \Sigma_{y^p} \overline{y}^p + \sum \limits_{m=2}^{n_l} \Sigma_{u^m} \overline{u}^m
\end{align}
%-------------------------------------------------------------
where term clusters and coefficients are defined as follows: constant terms in $\Sigma_0$; linear terms in $y$, $\Sigma_y \overline{y}$; linear terms in $u$, $\Sigma_u \overline{u}$; cross-terms in $\sum \limits_{m=1}^{n_l-1} \sum \limits_{p=1}^{n_l-m} \Sigma_{y^p u^m} \overline{y}^p \ \overline{u}^m$; non-linear terms in $y$, $\sum \limits_{p=2}^{n_l} \Sigma_{y^p} \overline{y}^p$; non-linear terms in $u$,$\sum \limits_{m=2}^{n_l} \Sigma_{u^m} \overline{u}^m$.

%-------------------------------------------------------------
\subsection{The Structure Selection Problem}
\label{s:MSS}

The identification of a system includes the following two steps: 1) Determination of a significant/system terms 2) Estimation of corresponding coefficients. Due to convenient linear-in-parameter form of the NARX models, the parameters can be estimated relatively easily with least-square based approaches. In contrast, detection of significant terms is a comparatively challenging task and it is often referred to as the \textit{structure selection problem}. This problem has been extensively studied for continuous, discrete and time-varying systems both in time and frequency domain~\cite{Billings:2013,Hafiz:Swain:2018b,Hafiz:Swain:2DUPSO:2019,Hafiz:Swain:Floating:2019,Hafiz:Swain:MOEA:2020}.

To understand the structure selection problem, consider the identification of a nonlinear system represented by polynomial NARX model. Given a large model set with $n$ number of terms, denoted as,
%---------------------------------------------------------
\begin{equation}
    \mathcal{X}_{model}=\begin{bmatrix} x_1 & x_2 & \dots & x_{n} \end{bmatrix}
\end{equation}
%---------------------------------------------------------
where, $x_1, x_2, \dots x_{n}$ represent any possible linear or non-linear term of the NARX model. The goal of the structure selection is to determine the \textit{optimum} subset of terms, $\mathcal{X}^{\star} \subset \mathcal{X}_{model}$, by minimizing a suitable criterion function,`$J(\cdotp)$'.

It is worth noting that the model set $\mathcal{X}_{model}$ is essentially the union of all the possible term clusters~\cite{Aguirre:Billings:1995b}, \textit{i.e.},
%---------------------------------------------------------
\begin{align}
    \mathcal{X}_{model} = & \bigcup \limits_{m=0\dots l;p=0 \dots m} \Omega_{y{^p}u^{m-p}} \\
    = & \{ \Omega_0 \cup \Omega_u \cup \Omega_y \cup \Omega_{y^2} \cup \Omega_{yu} \cup \Omega_{u^2} \cup \dots  \}
\end{align}
%---------------------------------------------------------
where, $\Omega_0$ denotes the constant term.
%-------------------------------------------------------------
\subsection{Pareto Dominance}
\label{s:PD}

It is often difficult to identify the optimal solution for multi-criteria/objective problems due to the contradictory nature of search objectives. In practice, the unique optimal solution to such problem may not exist, in contrast, there exist multiple solutions which are \textit{non-dominated} or \textit{Pareto Optimal}, \textit{i.e.}, the solutions which are not necessarily optimum for each objective however better than the other solutions when all objectives are simultaneously considered.

To understand the concept of Pareto dominance, consider two structures $\mathcal{X}_1$ and $\mathcal{X}_2$ with the corresponding criteria/objectives, as follows:
% ---------------------------------------------------------
\begin{align*}
 \vec{J}(\mathcal{X}_1) & =\begin{Bmatrix} J_1(\mathcal{X}_1), & J_2(\mathcal{X}_1), & \dots & J_{n_{obj}}(\mathcal{X}_1) \end{Bmatrix}\\
 \text{and, \ } \vec{J}(\mathcal{X}_2) & =\begin{Bmatrix} J_1(\mathcal{X}_2), & J_2(\mathcal{X}_2), & \dots & J_{n_{obj}}(\mathcal{X}_2) \end{Bmatrix}
\end{align*}
% ---------------------------------------------------------
where, `$n_{obj}$' denotes the number of search objectives.

The Pareto dominance for these structures can be determined on the basis of the objective values as follows: $\mathcal{X}_1$ dominates $\mathcal{X}_2$,
%----------------------------------------------------------------------
\begin{align}
  if{f} \ \ &  \forall{p} \in \{1 \dots n_{obj}\} : \, J_p(\mathcal{X}_1) \leq J_p(\mathcal{X}_2) \nonumber\\ 
  & \wedge \: \exists{p} \in \{1 \dots n_{obj}\} : J_p(\mathcal{X}_1) < J_p(\mathcal{X}_2)
\end{align}
%----------------------------------------------------------------------
This is denoted by $\mathcal{X}_1 \prec \mathcal{X}_2$.

%-------------------------------------------------------------
\section{Proposed Grey-Box Identification Approach}
\label{s:propsedapproach}

The main objective of this study is to identify a model which can yield a better dynamic prediction as well as provide a valid static behavior of a buck converter over a wide input range. It has been shown that \textit{a priori} information about the static behavior of the buck converter can be integrated into the structure selection process, albeit with a trade-off in dynamic prediction capability~\cite{Aguirre:Donoso:2000,Correa:Aguirre:2002}. The proposed approach, therefore, casts the grey-box identification problem into a multi-objective framework to obtain a better overall trade-off over the desired objectives. In this approach, both dynamic prediction capability and static behavior, are explicitly formulated as the search objectives and integrated into the multi-objective structure selection procedure.

In particular, this study takes a two-pronged approach to exploit the \textit{a priori} information about the static behavior. First, the static behavior is used to determine the set of viable term-clusters. This step leads to a significant reduction in the search space by removing non-essential clusters as will be discussed in Section~\ref{s:useofprior}. Next, the static function of the model under consideration is determined and compared with the known static behavior. This quantification of the static behavior is the key feature of the proposed approach where this is explicitly included as one of the search objectives. 

% to include static performance as one of the search objectives in the proposed approach. This is discussed in Section~\ref{s:MOSS}. 

%-------------------------------------------------------------
\subsection{Prior Knowledge}
\label{s:useofprior}

Given that the static input-output relation of the buck converter is known, it can be used to identify the viable term clusters. To this end, the static behavior in~(\ref{eq:static}) can be represented in a polynomial form as follows:
%-------------------------------------------------------------
\begin{align}
    \label{eq:static2}
    \overline{y} & = b_0 + b_1 \overline{u} \\
    \text{where, } b_0 & =  \frac{4V_d}{3}, \quad b_1 = - \frac{V_d}{3} \nonumber
\end{align}
%-------------------------------------------------------------
It is, thus, clear that to induce such a static behavior in the identified model, the corresponding static function should be a polynomial of input, $u$. 
Further, the static function of the NARX model can be determined from~(\ref{eq:cluster3}) as follows:
%-------------------------------------------------------------
\begin{align*}
% \label{eq:clusterNarx}
    \overline{y} = & \frac{\Sigma_0 + \Sigma_u \overline{u} + \sum \limits_{m=2}^{n_l} \Sigma_{u^m} \overline{u}^m}{1-\Sigma_y - \sum \limits_{m=1}^{n_l-1} \sum \limits_{p=1}^{n_l-m} \Sigma_{y^p u^m} \overline{y}^{p-1} \ \overline{u}^m - \sum \limits_{p=2}^{n_l} \Sigma_{y^p} \overline{y}^{p-1}}
\end{align*}
%-------------------------------------------------------------

It is easy to see that the following conditions should be satisfied in order to induce the static behavior similar to~(\ref{eq:static2}),
%-------------------------------------------------------------
\begin{align}
\label{eq:cond}
\Sigma_{y^p u^m} & = 0, \quad m=1,\dots, n_l-1, \text{ and } p=1,\dots, n_l-m \nonumber\\
\Sigma_{y^p} & = 0, \quad p=2, \dots n_l
\end{align}
%-------------------------------------------------------------
which yields, 
%-------------------------------------------------------------
\begin{align}
\label{eq:cluster4}
     \overline{y} = & \frac{\Sigma_0 + \Sigma_u \overline{u} + \sum \limits_{m=2}^{n_l} \Sigma_{u^m} \overline{u}^m}{1-\Sigma_y}
\end{align}
%-------------------------------------------------------------
This can further be simplified as,
%-------------------------------------------------------------
\begin{align}
\label{eq:clusterPoly}
     \overline{y} = & a_0 + a_1 \overline{u} + a_2 \overline{u}^2 + \dots + a_{n_l} \overline{u}^{n_l}\\
     \text{where, } a_0= & \frac{\Sigma_0}{1-\Sigma_y}, a_1=\frac{\Sigma_u}{1-\Sigma_y}, \dots, a_{n_l}=\frac{\Sigma_{u^{n_l}}}{1-\Sigma_y} \nonumber
\end{align}
%-------------------------------------------------------------
The static relation is now in the desired polynomial form. Further, the required conditions for this simplification~(\ref{eq:cond}), can easily be satisfied by excluding the terms from the non-linear output cluster and the cross-term clusters from the pool of candidate terms, \textit{i.e.}, 
%-------------------------------------------------------------
\begin{align}
\mathcal{X}_{model} & =\mathcal{X}_{model} \setminus \Omega_{y^p}, \quad p=2, \dots,n_l\\
\mathcal{X}_{model} & =\mathcal{X}_{model} \setminus \Omega_{y^p u^m},  \nonumber\\ 
\text{where, } m & =1, \dots, n_l-1; \text{ and } p=1, \dots n_l-m \nonumber
\end{align}
%-------------------------------------------------------------

It is worth emphasizing that although this reduction in candidate terms is crucial to induce the desired static behavior, it often involves a trade-off in the dynamic prediction capabilities~\cite{Aguirre:Donoso:2000}. Hence, although further reduction in pool of candidate term is possible, it is not desirable. This issue is discussed through an illustrative example in Section~\ref{s:CommTC}.

%-------------------------------------------------------------
\subsection{Multi-objective Structure Selection}
\label{s:MOSS}

The structure selection is inherently multi-objective in nature as it involves the following two decisions: 1) \textit{How many terms are required to represent the system dynamics?} and 2) \textit{Which are the significant terms among candidate terms?} These two issues are crucial to effectively address the \textit{bias-variance} dilemma. Hence, the structure selection can be approached as the multi-objective optimization problem. Further, the criterion function to evaluate a subset of candidate terms or \textit{structure} can be formulated as follows:
% ---------------------------------------------------------
\begin{align}
    \label{eq:mss1}
    \arg \min \ \vec{J}(\mathcal{X}_i) & = \begin{bmatrix} J_1(\mathcal{X}_i), &  J_2(\mathcal{X}_i) \end{bmatrix}\\
    \text{where, } J_1(\mathcal{X}_i) & = \xi_i, \quad J_2(\mathcal{X}_i) = \mathcal{E}_i \nonumber
\end{align}
% ---------------------------------------------------------
`$\mathcal{X}_i$' denotes the $i^{th}$ structure under consideration; `$\xi_{i}$' denotes the cardinality (\textit{number of terms}) in $\mathcal{X}_i$; `$\mathcal{E}_i$' denotes the free-run prediction error obtained over the validation data and it is given by,
% ---------------------------------------------------------
\begin{align}
    \label{eq:errorpredict}
    \mathcal{E}_i & = \frac{1}{\mathcal{N}_v} \sum \limits_{k=1}^{\mathcal{N}_v} [ y(k) - \hat{y}(k) ]^2
\end{align}
% ---------------------------------------------------------
where, `$\hat{y}$' denotes the model predicted (\textit{free run} or \textit{simulated}) output obtained with $\mathcal{X}_i$; and `$\mathcal{N}_v$' denotes the length of the validation data.

It is worth noting that since the criterion function in (\ref{eq:mss1}) directly incorporates the free-run prediction error ($\mathcal{E}$) and the structure cardinality ($\xi_i$), the search process to optimize $\vec{J}(\cdot)$ is likely to yield parsimonious models with a better dynamic prediction capability. Similarly, if somehow the static behavior can be quantified and explicitly formulated as one of the search objectives then the search can be directed to identify the models with all the desired `\textit{qualities}', \textit{i.e.}, compact models with better dynamic prediction and globally valid static behavior. This has been the main motivation for the proposed approach.

Given that the static behavior of the buck converter is known (given by~(\ref{eq:static})) and the same for a candidate model can be determined using~(\ref{eq:cluster4}), the static behavior can easily be quantified for the search purposes as follows: 
% ---------------------------------------------------------
\begin{align}
    \label{eq:estatic}
    \overline{\mathcal{E}}_i & = \sum \limits_{k=1}^{\mathcal{N}_s} [ \overline{y}_{buck}(k) - \overline{y}(k) ]^2 %\frac{1}{\mathcal{N}_s}
\end{align}
% ---------------------------------------------------------
where, `$\mathcal{N}_s$' denotes the length of the static validation data; `$\overline{y}_{buck}$' denotes the steady state output of buck converter which is given by~(\ref{eq:static}); `$\overline{y}$' denotes the steady state output of the $i^{th}$ structure $\mathcal{X}_i$. 

It is worth noting that the static behavior can still be quantified even when the explicit input-output static relation similar to~(\ref{eq:static}) is not available. For such a scenario, the required static data could be obtained experimentally, by the steady-state input-output measurements.

Since the static behavior of the candidate structure can be quantified using~(\ref{eq:estatic}), it is now possible to integrate static behavior as one of the search objectives, as follows:
% ---------------------------------------------------------
\begin{align}
    \label{eq:mss2}
    \arg \min \ \vec{J}(\mathcal{X}_i) & = \begin{bmatrix} J_1(\mathcal{X}_i), &  J_2(\mathcal{X}_i), & J_3(\mathcal{X}_i) \end{bmatrix}\\
    \text{where, } J_1(\mathcal{X}_i) & = \xi_i, \quad J_2(\mathcal{X}_i) = \mathcal{E}_i, \quad J_3(\mathcal{X}_i) = \overline{\mathcal{E}}_i.\nonumber
\end{align}
% ---------------------------------------------------------
%-----------------------------------------------------------------
%------                     Pseudo Code: MOEA - common
%-----------------------------------------------------------------
\begin{algorithm}[!t]
    \small
    \SetKwInOut{Input}{Input}
    \SetKwInOut{Output}{Output}
    \SetKwComment{Comment}{*/ \ \ \ }{}
    \Input{Population/Archive of `$ps$' parents, $\beta_1, \ \beta_2, \ \dots, \ \beta_{ps}$}
    \Output{Population of `$ps$' offspring, $\hat{\beta}_1, \ \hat{\beta}_2, \ \dots, \ \hat{\beta}_{ps}$}
    \algorithmfootnote{`CTS' denotes the \textit{Crowded Tournament Selection}, see~\cite{Deb:Pratap:2002};\\ \smallskip `BTS' denotes the \textit{Binary Tournament Selection}, see~\cite{Zitzler:Laumanns:2001}}
    \BlankLine
    \Comment*[h] {Selection \& Crossover}\\
    \For{i = 1 to $\frac{ps}{2}$}
    {

        \BlankLine
        \Comment*[h] {NSGA-II: Crowded Tournament Selection}\\
        \BlankLine
        $\{ \beta_p, \ \beta_q \} = CTS$ (population, \ ranks, \ crowding \ distance)
        \BlankLine
        \BlankLine
        \Comment*[h] {SPEA-II: Binary Tournament Selection}\\
        \BlankLine
        $\{ \beta_p, \ \beta_q \} = BTS$ (archive, \ pseudo \ fitness)
        \BlankLine
        \BlankLine
        \Comment*[h] {Parameterized Uniform Crossover}\\
        \BlankLine
        $\hat{\beta}_p \leftarrow \beta_p$, $\hat{\beta}_{q} \leftarrow \beta_q$\\ 
        \BlankLine
        \If{$p_c>rand$}
        {
         \BlankLine
         \For{j = 1 to $n$}
         {
                \BlankLine
                \If{$0.5>rand$}
                    {\BlankLine
                     $\hat{\beta}_{p,j} \leftarrow \beta_{q,j}, \quad \hat{\beta}_{q,j} \leftarrow \beta_{p,j}$}
         } % End of for loop
        }% End of pc If
    } % End of ps/2 for loop
    \BlankLine
    \BlankLine
    \Comment*[h] {Mutation}\\
    \For{i = 1 to $ps$}
    {
       \BlankLine
       \For{j = 1 to $n$}
        {
         \BlankLine
        \If{$p_m>rand$} 
         {
           $\hat{\beta}_{i,j} = 1-\hat{\beta}_{i,j}$ 
         } % End of pm If
        }% End of for loop
       Evaluate the fitness of $\hat{\beta}_{i}$ as per Algorithm~\ref{al:crf}
    } % End of mutation loop
   
\caption{Reproduction procedures}
\label{al:moea}
\end{algorithm}
%-----------------------------------------------------------------

Note that essentially this is a combinatorial optimization problem. An exhaustive search of all possible term subsets to solve~(\ref{eq:mss2}) is often intractable even for a moderate number of NARX terms `$n$', as it requires the examination of $2^{n}$ term subsets/structures. Hence, it is clear that an effective search strategy is crucial to optimize the multi-objective structure selection problem given by~(\ref{eq:mss2}). This can be accomplished by any multi-objective evolutionary algorithm such as NSGA-II~\cite{Deb:Pratap:2002}, SPEA-II~\cite{Zitzler:Laumanns:2001}, MOEA/D~\cite{Zhang:Li:2007} and others. The comparative analysis of these algorithms on the structure selection problem in~\cite{Hafiz:Swain:MOEA:2020} indicates that \textit{dominance} based MOEAs (\textit{e.g.}, NSGA-II and SPEA-II) often yields an improved Pareto front in comparison to \textit{decomposition} based MOEAs such as MOEA/D. Hence, in this study, NSGA-II and SPEA-II are selected to solve the structure selection problem given in~(\ref{eq:mss2}).

%---------------------------------------------------------------------
%------                     Pseudo Code:Criterion Function
%---------------------------------------------------------------------
\begin{algorithm}[!t]
    \small
    \SetKwInOut{Input}{Input}
    \SetKwInOut{Output}{Output}
    \SetKwComment{Comment}{*/ \ \ \ }{}
    \Input{Search Agent, $\beta_i$}
    \Output{$\vec{J}(\mathcal{X}_i)=\begin{Bmatrix} J_1(\mathcal{X}_i) & J_2(\mathcal{X}_i) & J_3(\mathcal{X}_i) \end{Bmatrix}$}
    % \Output{Criterion Function,$\vec{J}(\mathcal{X}_i)=\{ J_1(\mathcal{X}_i), J_2(\mathcal{X}_i), J_3(\mathcal{X}_i)\}$}
    %$\vec{J}(\mathcal{X}_i)=\begin{Bmatrix} J_1(\mathcal{X}_i) & J_2(\mathcal{X}_i) & J_3(\mathcal{X}_i) \end{Bmatrix}$
    \BlankLine
    Set the $i^{th}$ structure to null vector, \textit{i.e.},
    $\mathcal{X}_i \leftarrow \varnothing$ and $\xi_i \leftarrow 0$\\
    
    \BlankLine
    \Comment*[h] {Decode the Parent}\\\nllabel{line:crf1}
    \BlankLine
    \For{m = 1 to n} 
        { \BlankLine
           \If{$\beta_{i,m}=1$}
           {\BlankLine
             $\mathcal{X}_i \leftarrow \{ \mathcal{X}_i \cup x_m \}$  \Comment*[h] {add the $m^{th}$ term}\\
             $\xi_i \leftarrow \xi_i+1$
            \BlankLine
           }
          \BlankLine
        } 
    \BlankLine\nllabel{line:crf2}
    \Comment*[h] {Parameter Estimation}\\
    Estimate Coefficients, `$\Theta$', corresponding to the terms in $\mathcal{X}_i$ using Least Squares based algorithm (see~\cite{Billings:2013})
    \BlankLine
    \BlankLine
    \Comment*[h] {Evaluate the Criterion Function}\\
    \BlankLine
    Determine the dynamic prediction error $\mathcal{E}_i$ using~(\ref{eq:errorpredict})\\
    \BlankLine
    Determine the error in the static behavior $\overline{\mathcal{E}}_i$ as per~(\ref{eq:estatic})\\
    \BlankLine
    $J_1(\mathcal{X}_i) \leftarrow \xi_i, \ \ J_2(\mathcal{X}_i) \leftarrow \mathcal{E}_i, \ \ J_3(\mathcal{X}_i) \leftarrow \overline{\mathcal{E}}_i$
\caption{Evaluation of Criterion Function, $\vec{J}(\cdotp)$}
\label{al:crf}
\end{algorithm}
%---------------------------------------------------------------------

To address the structure selection problem with $n-$number of NARX terms, each \textit{parent} in MOEA encodes a candidate structure in an $n-$dimensional binary vector as follows:
%---------------------------------------------------------
\begin{align}
    \beta_i & = \begin{bmatrix} \beta_{i,1} & \beta_{i,2} & \dots & \beta_{i,n} \end{bmatrix}\\
    \text{where, \ } & \beta_{i,m} \in \{0,1\}, \ \ m=1,2,\dots n\nonumber
\end{align}
%---------------------------------------------------------
where, the $i^{th}$ parent, $\beta_i$, encodes $i^{th}$ structure $\mathcal{X}_i$. The $m^{th}$ term ($x_m$) from $\mathcal{X}_{model}$ is included into the candidate structure $\mathcal{X}_i$ provided the corresponding bit in the parent, `$\beta_{i,m}$' is set to `$1$'. For more details see the illustrative example in Appendix~\ref{s:aSr}.

Drawing on the recommendations in~\cite{Hafiz:Swain:MOEA:2020}, the \textit{qualitative} and \textit{quantitative} control parameters of MOEAs are set as follows:
\begin{itemize}
    \item NSGA-II: \textit{Population Size}: $50$; \textit{Selection}: crowded tournament selection~\cite{Deb:Pratap:2002}; \textit{Recombination}: uniform crossover; \textit{crossover rate} ($p_c$): $0.9$; \textit{Mutation}: flip-bit mutation; and \textit{mutation rate} ($p_m$): $0.006$.
    
    \item SPEA-II: \textit{Population Size}: $50$; \textit{Selection}: binary tournament selection~\cite{Zitzler:Laumanns:2001}; \textit{Recombination}: uniform crossover; \textit{crossover rate} ($p_c$): $0.7$; \textit{Mutation}: flip-bit mutation; and \textit{mutation rate} ($p_m$): $0.008$.
\end{itemize} 

 The reproduction operators being used in this study are shown in Algorithm~\ref{al:moea}, where `$\beta_i$' and `$\hat{\beta}_i$' respectively denote the $i^{th}$ parent and the corresponding offspring. Each parent under consideration is evaluated following the steps outlined in Algorithm~\ref{al:crf}. Note that the other search components of NSGA-II and SPEA-II such as non-dominated sorting, crowding distance and pseudo fitness are omitted here for sake of brevity. Further implementation details about MOEAs can be found in~\cite{Deb:Pratap:2002,Zitzler:Laumanns:2001,Hafiz:Swain:MOEA:2020}. 

The overall procedures involved in the proposed approach are outlined in Algorithm~\ref{al:moss}. Because of the stochastic nature of the algorithm, `$R$' independent runs are carried out. Each run is set to terminate after $25,000$ Function Evaluations (FEs). In each run, non-dominated structures and the corresponding criterion function are respectively accumulated in $\Gamma$ and $\Lambda$, as outlined in Line~\ref{line:moss1}-\ref{line:moss2}, Algorithm~\ref{al:moss}. At the end of these runs, the dominance of the accumulated structures in $\Gamma$ is again determined and the non-dominated structures and the corresponding criterion functions are stored respectively in $\Gamma^*$ and $\Lambda^*$. 

It is clear that the identified non-dominated structures in $\Gamma^*$ essentially represent a trade-off of varying degree over the search objectives, hence the \textit{a posteriori} selection of a particular structure from this pool is primarily dependent on the choice of the Decision Maker (DM). These issues are discussed in detail in the following subsection.
%---------------------------------------------------------------------
%------                     Pseudo Code:Proposed Approach
%---------------------------------------------------------------------
% \setlength{\textfloatsep}{2pt}% Remove \textfloatsep
\begin{algorithm}[!t]
    \small
    \SetKwInOut{Input}{Input}
    \SetKwInOut{Output}{Output}
    \SetKwComment{Comment}{*/ \ \ \ }{}
    \Input{Input-output Data, $(u,y)$}
    \Output{Identified Non-dominated Structures, $\Gamma^*$}
    \BlankLine
    % \Comment*[h] {Data Pre-processing}\\
    % \BlankLine
    %Decimate or filter the identification data if required\\
    % Split the input-output data into the \textit{estimation} and \textit{validation} sets\\
    Generate set of candidate NARX terms $\mathcal{X}_{model}$ as per~(\ref{eq:NARXmodel})\\ %by specifying $n_u, n_y$ and $n_l$ \\
    \BlankLine
    Remove all the terms in the nonlinear output clusters, \textit{i.e.}, $\mathcal{X}_{model}=\{ \mathcal{X}_{model} \setminus \Omega_{y^p}, p=2, \dots n_l \}$ \nllabel{line:clr1}\\
    \BlankLine
    Remove all the terms in the input-output cross-term clusters, \textit{i.e.}, $\mathcal{X}_{model}=\{ \mathcal{X}_{model} \setminus \Omega_{y^p u^m}, m=1, \dots n_l - 1, p=1, \dots n_l - q \}$ \nllabel{line:clr2}
    \BlankLine
    \BlankLine
    \Comment*[h] {Search for non-dominated structures}\\
    \BlankLine
    $\Gamma \leftarrow \varnothing$, $\Lambda \leftarrow \varnothing$
    \BlankLine
    Perform $R$ independent runs of MOEA \nllabel{line:1}\\
    \BlankLine
    \For{k = 1 to $R$} 
        { \BlankLine
          Record the non-dominated structures, \textit{i.e.},\\
          $\Gamma \leftarrow \Gamma \cup \begin{Bmatrix} \mathcal{X}_1 & \mathcal{X}_2 & \dots \end{Bmatrix}$  \nllabel{line:moss1}\\ 
          $\Lambda \leftarrow \Lambda \cup \begin{Bmatrix} \vec{J}(\mathcal{X}_1) & \vec{J}(\mathcal{X}_2) & \dots \end{Bmatrix}$\nllabel{line:moss2}
          \BlankLine
        } 
     \BlankLine
     Keep only the non-dominated structures, \textit{i.e.}, $\Gamma^* \prec \Gamma, \ \Lambda^* \prec \Lambda$\\
     \BlankLine
     \BlankLine
     \Comment*[h] {A posteriori Selection}\\
     \BlankLine
     Select a structure following MMD approach (see Algorithm~\ref{al:mmd})\\
     \BlankLine
     Select a structure following MTD approach (see Algorithm~\ref{al:mtd})
     \BlankLine
\caption{Proposed Grey-Box Identification}
\label{al:moss}
\end{algorithm}
%---------------------------------------------------------------------
%---------------------------------------------------------------------
\subsection{Preference Articulation}
\label{s:PrA}

The \textit{a posteriori} selection from the identified non-dominated structures in $\Gamma^*$ is primarily dependent on the choice of the Decision Maker (DM). To this end, two possible \textit{a posteriori} scenarios are considered in this study: 1) DM is unbiased, \textit{i.e.}, an equal preference is given to each design objective. 2) DM is biased towards a particular search objective. In the following, two \textit{a posteriori} solution selection techniques are briefly discussed which can accommodate these two distinct scenarios. 
%-----------------------------------------------------------------
\subsubsection{Minimum Manhattan Distance}
\label{s:MMD}

The Minimum Manhattan Distance (MMD)~\cite{Chiu:Yen:2016} approach for \textit{a posteriori} decision making is appropriate when an equal priority is assigned to each objective, \textit{i.e.}, the DM is unbiased. In this approach, the identified non-dominated structures in $\Gamma^*$ are ranked as follows: First, a hypothetical ideal point ($\vec{J}^{\star}$), which consists of the best value of each objective in $\Lambda^*$, is located in the objective space: 
%--------------------------------------------------------------------
\begin{equation}
    \vec{J}^{\star} = \begin{Bmatrix} J_1^{min}, & J_2^{min}, & \dots & J_{n_{obj}}^{min} \end{Bmatrix}
\end{equation}
%--------------------------------------------------------------------
where, $J_p^{min} = \min  J_p(\mathcal{X}_i) , \ \forall{\mathcal{X}_i} \in \Gamma^*  \ \text{and, \ } p=1, \dots, n_{obj}$.

Subsequently, for each non-dominated structure $\mathcal{X}_i \in \Gamma^*$, the Manhattan distance, $\mathcal{D}(\cdot)$, is evaluated with respect to $\vec{J}^\star$, as outlined in Line \ref{line:mmd1}-\ref{line:mmd2}, Algorithm~\ref{al:mmd}. Note that the Manhattan distance $\mathcal{D}(\cdot)$ is determined in the normalized objective space to avoid scaling issues. In the final step, the solutions are ranked in the ascending order of $\mathcal{D}(\cdot)$. Based on this ranking, a few top structures can be selected for further analysis to account for uncertainties associated with the measurement of the dynamical and the static data. However, in this study, only the structure corresponding to the minimum Manhattan distance, $\mathcal{D}(\cdot)$, is selected for sake of brevity.
%-----------------------------------------------------------------
%------                     Pseudo Code: MMD
%-----------------------------------------------------------------
\begin{algorithm}[!t]
    \small
    \SetKwInOut{Input}{Input}
    \SetKwInOut{Output}{Output}
    \SetKwComment{Comment}{*/ \ \ \ }{}
    \Input{Pareto set, $\Gamma^*=\{ \mathcal{X}_1, \mathcal{X}_2, \dots \}$\\
    Pareto front, $\Lambda^*=\{ \vec{J}(\mathcal{X}_1), \vec{J}(\mathcal{X}_2), \dots \}$}
    \Output{Selected Structure, $\mathcal{X}^{*}$}
    \BlankLine
    \Comment*[h] {`Ideal' and `worst' Points}\\
    \BlankLine
    % $\vec{J}^{\star}\leftarrow \varnothing$\\
    \For{i = 1 to $n_{obj}$}
    {
        \BlankLine
        $J_p^{min} = \arg \min  J_p(\mathcal{X}_i) , \ \forall{\mathcal{X}_i} \in \Gamma^*$\\
        \BlankLine
        $J_p^{max} = \arg \max  J_p(\mathcal{X}_i) , \ \forall{\mathcal{X}_i} \in \Gamma^*$\\
        % $\vec{J}^{\star}\leftarrow \vec{J}^{\star} \cup J_p^{min}$
    }
    \BlankLine
    \BlankLine
    \Comment*[h] {Distance Evaluation}\\
    \For{j = 1 to $|\Gamma^*|$ \nllabel{line:mmd1}}
    {
       \BlankLine
       \For{p = 1 to $n_{obj}$}
       {
            \BlankLine
            $d_p(\mathcal{X}_j,J_p^{min}) =  \displaystyle \abs{\frac{J_p(\mathcal{X}_j)- J_p^{min}}{J_p^{max}- J_p^{min}}}$
       }
       \BlankLine
       Determine the Manhattan distance metric, 
       $\mathcal{D}(\mathcal{X}_j) = \displaystyle \sum \limits_{p=1}^{n_{obj}} d_p(\mathcal{X}_j,J_p^{min})$
    }\nllabel{line:mmd2} % End of j loop
   \BlankLine
    Select the structure with the minimum distance, \textit{i.e.}, $\mathcal{X}^{*} = \{ \mathcal{X}_i | \mathcal{D}(\mathcal{X}_i) = \arg \min \mathcal{D}(\mathcal{X}_k), \forall{\mathcal{X}_k} \in \Gamma^* \}$%\limits_{\forall{\mathcal{X}_k} \in \Gamma^*}
    \BlankLine
\caption{MMD approach to \textit{a posteriori} selection}
\label{al:mmd}
\end{algorithm}
%-----------------------------------------------------------------
%-----------------------------------------------------------------
%------                     Pseudo Code: MTD
%-----------------------------------------------------------------
\begin{algorithm}[!t]
    \small
    \SetKwInOut{Input}{Input}
    \SetKwInOut{Output}{Output}
    \SetKwComment{Comment}{*/ \ \ \ }{}
    \Input{Pareto set, $\Gamma^*=\{ \mathcal{X}_1, \mathcal{X}_2, \dots \}$\\
    Pareto front, $\Lambda^*=\{ \vec{J}(\mathcal{X}_1), \vec{J}(\mathcal{X}_2), \dots \}$}
    \Output{Selected Structure, $\mathcal{X}^{*}$}
    \BlankLine
    \Comment*[h] {Preference formulation}\\
     \BlankLine
     Specify the objective rankings, $O = \begin{bmatrix} O_{\xi} & O_{\mathcal{E}} & O_{\overline{\mathcal{E}}} \end{bmatrix}$\\
     \BlankLine
     Select the preference intensity, $\mathcal{I} \in [1,9]$\\
    %  Determine the priority weights, $\vec{w}$\\
    \BlankLine
     \For{i = 1 to $n_{obj}$} 
     {  
         \BlankLine
         \For{j = 1 to $n_{obj}$ \nllabel{line:mtd1}}
         {
           \BlankLine
           $\delta_O = \frac{O_j - O_i}{n_{obj}-1}$\\
           \BlankLine
        %   \BlankLine
           $\tau_{i,j} = \mathcal{I}^{\delta_O}$ \Comment*[h] {preference relations}
         }\nllabel{line:mtd2} % end of j
         \BlankLine
         $w_i = \displaystyle \Big( \prod \limits_{j=1}^{n_{obj}} \tau_{i,j} \Big)^{1/n_{obj}}$ \nllabel{line:mtd3}
     }% end of i
    \BlankLine
    $\vec{w} = \displaystyle \frac{\begin{bmatrix} w_1 & w_2 & \dots & w_{n_{obj}} \end{bmatrix}}{\sum_{p=1}^{n_{obj}} w_p}$ \Comment*[h] {priority weights}
    \BlankLine
    \BlankLine
    \Comment*[h] {Tournament function}\\
    \BlankLine
    \For{i = 1 to $|\Gamma^*|$}
    {
        \BlankLine
        \For{p = 1 to $n_{obj}$ \nllabel{line:mtd5}} 
        {
           \BlankLine
           $t_{i,p} \leftarrow 0$\\
           \BlankLine
           \For{j = 1 to $|\Gamma^*|$}
           {
                \BlankLine
                \If{$J_{p}(\mathcal{X}_j)-J_{p}(\mathcal{X}_i)>0$}
                {
                    \BlankLine
                    $t_{i,p} \leftarrow t_{i,p} + 1$
                }
           }% End of j-loop
           $T_p(\mathcal{X}_i,\Gamma^*) = \displaystyle \frac{t_{i,p}}{|\Gamma^*|-1} $ %\displaystyle \sum \limits_{i=1}^{|\Gamma^*|}
        }\nllabel{line:mtd6} % End of p-loop
        \BlankLine
        Determine global rank, 
        $\mathcal{R}(\mathcal{X}_i) = \displaystyle \Big( \prod \limits_{p=1}^{n_{obj}} T_p(\mathcal{X}_i,\Gamma^*)^{w_p} \Big)^{1/n_{obj}}$ \nllabel{line:mtd4}
    }% End of i-loop
     \BlankLine
     Select the structure with the maximum global rank $\mathcal{R}(\cdot)$, \textit{i.e.}, $\mathcal{X}^{*} = \{ \mathcal{X}_i | \mathcal{R}(\mathcal{X}_i) = \arg \max \mathcal{R}(\mathcal{X}_k), \forall{\mathcal{X}_k} \in \Gamma^* \}$%\limits_{\forall{\mathcal{X}_k} \in \Gamma^*}
     \BlankLine
\caption{MTD approach to \textit{a posteriori} selection}
\label{al:mtd}
\end{algorithm}
%-----------------------------------------------------------------
%------------------------------------------------------------------
\begin{table}[!t]
  \centering
  \small
  \caption{Objective Rankings for \textit{a posteriori} selection with MTD} \label{t:ObjRank}%
  \begin{adjustbox}{max width=0.49\textwidth}
    \begin{tabular}{cc}
    \toprule
    \makecell{\textbf{Objective Rankings}\\ \boldmath$O = \begin{bmatrix} O_\xi & O_{\mathcal{E}} & O_{\overline{\mathcal{E}}} \end{bmatrix}$} & \makecell{\textbf{Weight Vector} \\ \boldmath$\vec{w} = \begin{bmatrix} w_\xi & w_{\mathcal{E}} & w_{\overline{\mathcal{E}}} \end{bmatrix}$}\\[0.8ex]
    \midrule
    $O_1 = \begin{bmatrix} 3 & 1 & 2 \end{bmatrix}$  & $\vec{w}_1 = \begin{bmatrix}0.1214 & 0.6071 & 0.2715\end{bmatrix}$ \\[1ex]
    $O_2 = \begin{bmatrix} 1 & 3 & 2 \end{bmatrix}$ & $\vec{w}_2 = \begin{bmatrix} 0.6071 & 0.1214 & 0.2715 \end{bmatrix}$ \\[1ex]
    $O_3 = \begin{bmatrix} 1 & 2 & 3\end{bmatrix}$ & $\vec{w}_3 = \begin{bmatrix} 0.6071 & 0.2715 & 0.1214 \end{bmatrix}$ \\[1ex]
    \bottomrule
    \end{tabular}%
  \end{adjustbox}
\end{table}%
%------------------------------------------------------------------
%-----------------------------------------------------------------
\subsubsection{Formulation of Priority Weights}
\label{s:PrW}

If the DM is biased towards a particular search objective, it is essential to embed such a preference in the \textit{a posteriori} selection. However, the human preferences are often abstract and partial~\cite{Branke:2008}, hence the first step is to encode such preferences in a quantitative metric. To this end, the DM's preferences are encoded into multiplicative preference relations following the approach proposed in~\cite{Zhang:Chen:2004}, as follows: First, the DM assigns a \textit{rank} (denoted by `$O$') to each objective in the order of preference. For example, if the parsimony and the static performance are preferred over the dynamic prediction, then the objective rankings are given by $\begin{bmatrix} O_\xi & O_{\mathcal{E}} & O_{\overline{\mathcal{E}}} \end{bmatrix} = \begin{bmatrix} 1 & 3 & 2 \end{bmatrix}$. 

Next, the intensity of the objective rankings, denoted by $\mathcal{I}$, is assigned on a scale from `$1$' to `$9$'. The preference intensity determines the strength of the specified objective rankings, \textit{e.g.}, $\mathcal{I}=1$ assigns equal importance to all the objectives whereas $\mathcal{I}=9$ denotes extreme prejudice. Based on the specified objective rankings ($O$) and the preference intensity ($\mathcal{I}$) the multiplicative preference relations (denoted by `$\tau$') are determined following the steps in Line \ref{line:mtd1}-\ref{line:mtd2}, Algorithm~\ref{al:mtd}. Here, `$\tau_{i,j}$' implies that the $i^{th}$ objective is $\tau_{i,j}$ times more important than the $j^{th}$ objective. Finally, the preference weights (denoted by `$w$') are determined as outlined in Line \ref{line:mtd3}, Algorithm~\ref{al:mtd}. This procedure is further explained through the illustrative example in Appendix~\ref{s:aPrw}.

It is worth noting that a total of $n_{obj}!$ combinations of objective rankings are possible for an $n_{obj}$ number of objectives. To highlight the effects of specified preferences, 3 distinct combinations of objective rankings are considered in this study. Further, the preference intensity is fixed to `$5$', \textit{i.e.}, $\mathcal{I}=5$. Table~\ref{t:ObjRank} gives the objective rankings and the corresponding weight vectors, which are being considered in this study.

%------------------------------------------------------------------
\subsubsection{Multi-criteria Tournament Decision}
\label{s:MTD}

Once the DM's preferences are quantified into the \textit{priority weights}, the next step is to embed these weights into the \textit{a posteriori} selection process. For this purpose, the Multi-criteria Tournament Decision (MTD) approach~\cite{Parreiras:Vasconcelos:2009} is considered, which ranks the identified non-dominated structures using the specified weights. In particular, for each structure $\mathcal{X}_i \in \Gamma^*$, the tournament function, is determined by a pairwise comparison with the remaining structures, as outlined in Line~\ref{line:mtd5}-\ref{line:mtd6}, Algorithm~\ref{al:mtd}.  The tournament function, denoted by $T_p(\mathcal{X}_i,\Gamma^*)$, essentially determines the total number of structures in $\Gamma^*$ compared to which $\mathcal{X}_i$ yields a better value for the $p^{th}$ objective. The similar procedure is repeated to determine this function for all the `$n_{obj}$' objectives. Finally, the \textit{global rank} for $\mathcal{X}_i$ across all objectives is determined by aggregating the tournament functions as outlined in Line~\ref{line:mtd4}, Algorithm~\ref{al:mtd}. This procedure is repeated to rank each structure $\mathcal{X}_i \in \Gamma^*$. The structure with the maximum global rank, $\mathcal{R}(\cdot)$, is selected as the final choice.

%%---- Pareto Front -----------------------------------------------
\begin{figure}[!t]
\small
\centering
\begin{subfigure}{0.37\textwidth}
  \includegraphics[width=\textwidth]{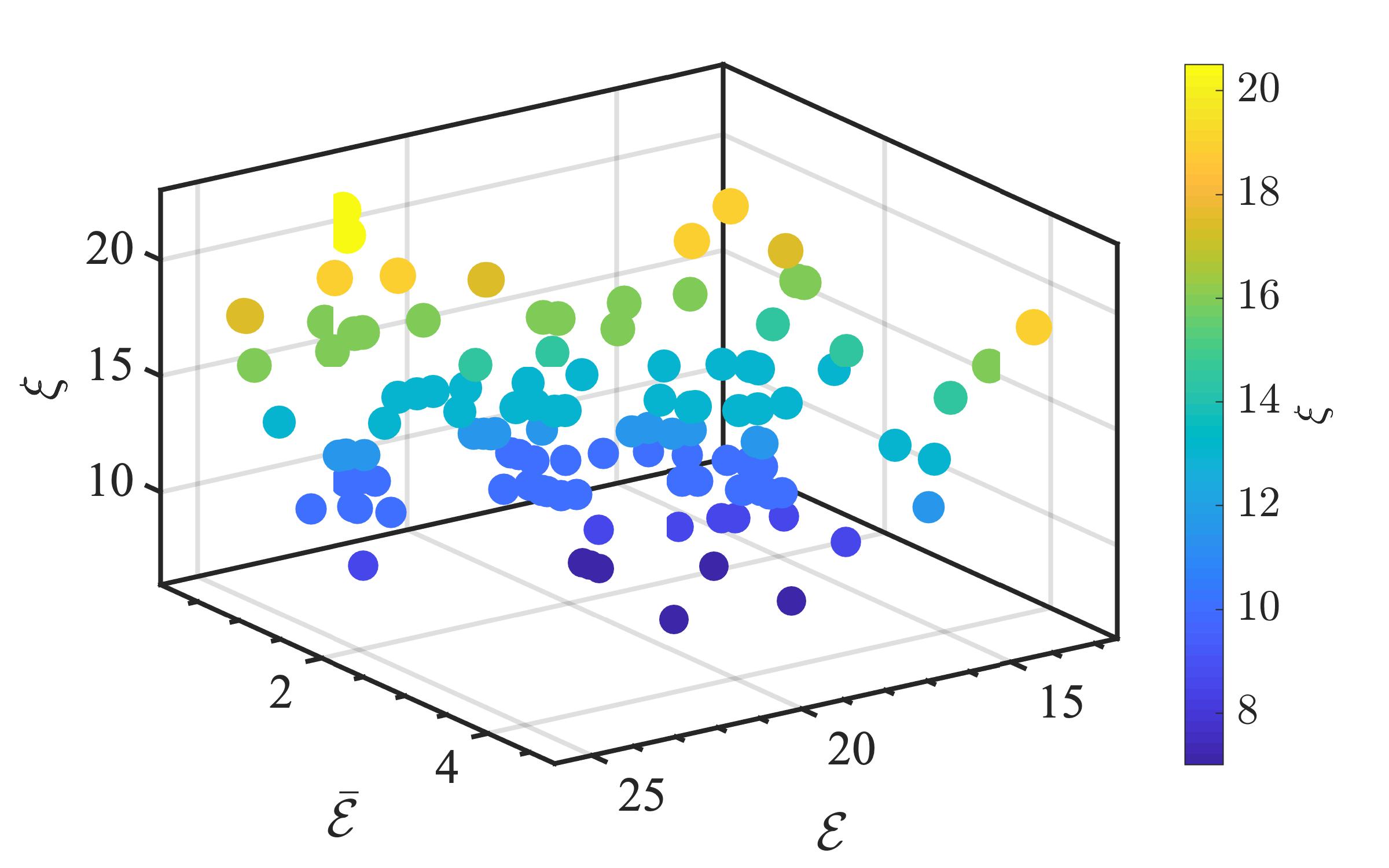}
  \caption{$\Lambda_A$, NSGA-II }
  \label{f:PF_A}
\end{subfigure}
\hfill
\begin{subfigure}{0.37\textwidth}
  \includegraphics[width=\textwidth]{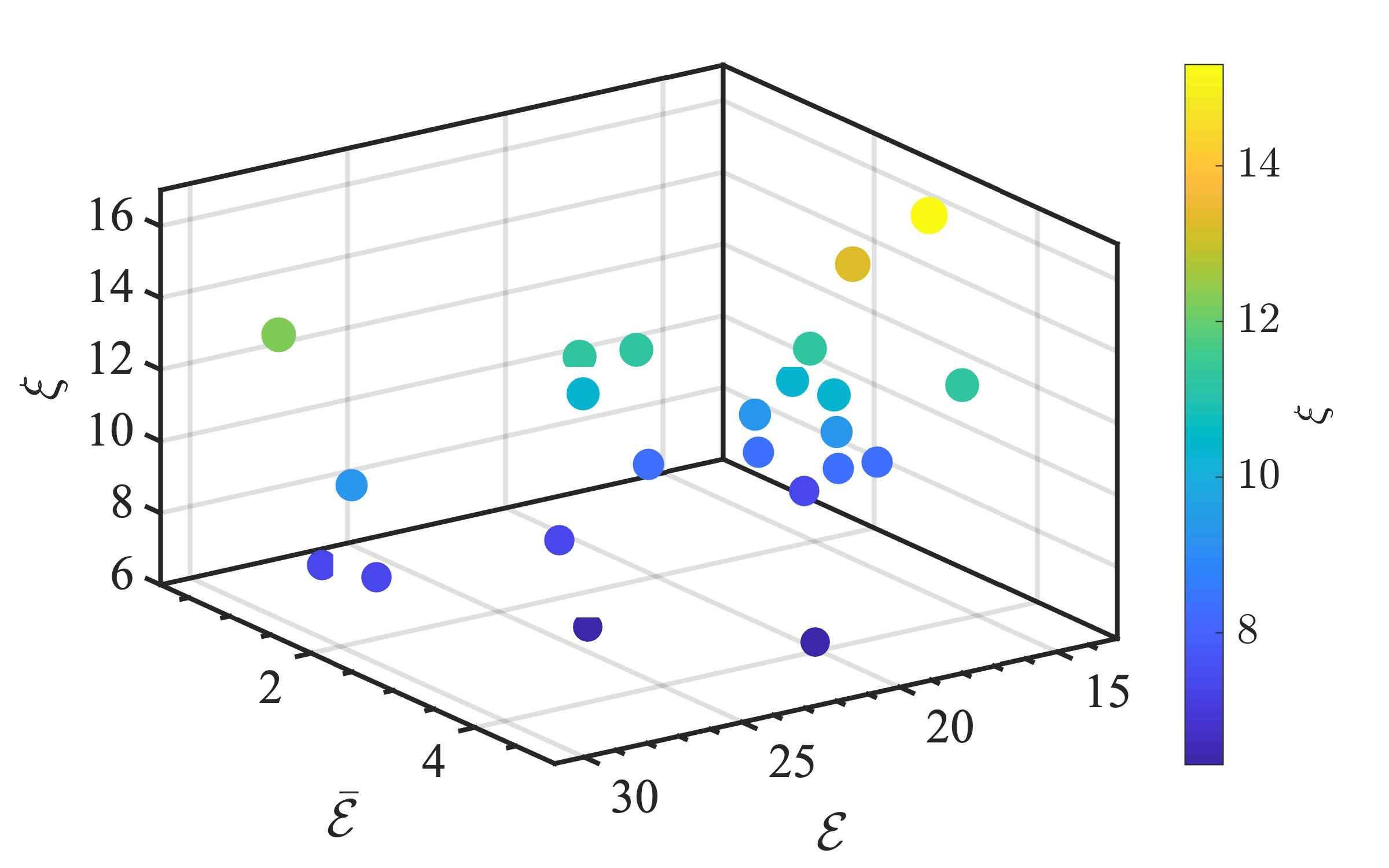}
  \caption{$\Lambda_B$, SPEA-II}
  \label{f:PF_B}
\end{subfigure}
\caption{Set of non-dominated structures found over 100 independent runs of MOEAs. $\xi$, $\mathcal{E}$ and $\overline{\mathcal{E}}$ are the search objectives and respectively denote the number of terms, the prediction error and the static error.}
\label{f:PF}
\end{figure}
%--------------------------------------------------------------------
%---------------------------------------------------------------------
\section{Results}
\label{s:res}

The goal of this study is to develop a new approach to embed \textit{a priori} system knowledge directly into the fundamental step of structure selection for grey-box identification problems. The efficacy of the proposed approach is demonstrated by considering a practical case study of buck converter modeling. The known static behavior of buck converter is treated as \textit{a priori} knowledge. In the following, the results of this case study are discussed in detail. First, the search behavior of MOEAs is compared in Section~\ref{s:res1}. The results of \textit{a posteriori} preference articulation are discussed next in Section~\ref{s:resPrfA}. The steady-state behavior of the identified models is determined in Section~\ref{s:SS}. Next, the results of a detailed comparative evaluation with the earlier investigation are provided in Section~\ref{s:res3}. Finally, the role of non-linear input clusters is discussed in Section~\ref{s:CommTC}.
%---------------------------------------------------------------------
\subsection{Search Outcome}
\label{s:res1}

The overall procedure followed to identify non-dominated structures is outlined in Algorithm~\ref{al:moss}. A total of $168$ data-points are obtained for identification purposes from the experimental buck converter setup described in Section~\ref{s:data}. Following the cross-validation principle, $100$ data points are used for the estimation of coefficients and the remaining data points form the validation data, \textit{i.e.}, $\mathcal{N}_v=68$. The candidate set of $286$ NARX terms (\textit{i.e.}, $n=286$) is generated by the following specifications of the NARX model in~(\ref{eq:NARXmodel}): $[n_u,n_y,n_l] = [5,5,3]$. Further, as discussed in Section~\ref{s:useofprior}, all the terms in nonlinear output and cross-term clusters are removed from the candidate terms, as outlined in Line~\ref{line:clr1}-\ref{line:clr2}, Algorithm~\ref{al:moss}. 

Following the steps outlined in Algorithm~\ref{al:moss}, a set of non-dominated structures are identified over $100$ independent runs of MOEAs. A total of $117$ and $24$ non-dominated structures are identified respectively by NSGA-II and SPEA-II. For sake of simplicity, let the set of non-dominated structures identified by NSGA-II and SPEA-II be denoted by `$\Gamma_A$' and `$\Gamma_B$', respectively. Similarly, denote the corresponding set of objective function vectors by `$\Lambda_A$' and `$\Lambda_B$', \textit{i.e.},

%---------------------------------------------------------------------
\begin{small}
\begin{align*}
    \Gamma_A & =\begin{Bmatrix} \mathcal{X}_1, \mathcal{X}_2, \dots, \mathcal{X}_{117} \end{Bmatrix}, \ \Lambda_A =\begin{Bmatrix} \vec{J}(\mathcal{X}_1), \vec{J}(\mathcal{X}_2), \dots, \vec{J}(\mathcal{X}_{117}) \end{Bmatrix}\\
    \Gamma_B & =\begin{Bmatrix} \mathcal{X}_1, \mathcal{X}_2, \dots, \mathcal{X}_{24} \end{Bmatrix}, \ \Lambda_B =\begin{Bmatrix} \vec{J}(\mathcal{X}_1), \vec{J}(\mathcal{X}_2), \dots, \vec{J}(\mathcal{X}_{24}) \end{Bmatrix}
\end{align*}
\end{small}
%---------------------------------------------------------------------

The approximate Pareto fronts obtained by NSGA-II ($\Lambda_A$) and SPEA-II ($\Lambda_B$) are shown in Fig.~\ref{f:PF_A} and~\ref{f:PF_B}, respectively. From these results, it is obvious that NSGA-II could identify significantly higher number of non-dominated structures, \textit{i.e.}, $|\Gamma_A|>|\Gamma_B|$. Further, the set coverage metric~\cite{Zitzler:Thiele:2003} is considered to compare the \textit{quality} of the identified structures. This metric is denoted here by  `$\mathcal{C}(\cdotp)$', and it is defined as:

%---------------------------------------------------------------------
\begin{small}
\begin{equation}
    \label{eq:coverageMetric}
    \mathcal{C}(A,B) = \frac{\Big| \Big\{ \mathcal{X}_{B,i} \in \Gamma_B | \ \exists{\mathcal{X}_{A,j}}\in \Gamma_A :  \mathcal{X}_{A,j} \preceq \mathcal{X}_{B,i}   \Big\}\Big|}{|\Gamma_B|}
\end{equation}
\end{small}
%---------------------------------------------------------------------

The metric $\mathcal{C}(A,B)$ essentially determines the number of structures in $\Gamma_B$ which are dominated by the structures in $\Gamma_A$. For the approximate Pareto fronts shown in Fig.~\ref{f:PF}, these metrics are determined to be: $\mathcal{C}(A,B)=0.6250$ and $\mathcal{C}(B,A)=0.0854$. It is easy to follow that $\mathcal{C}(A,B)>\mathcal{C}(B,A)$ which implies that the search performance of NSGA-II is better than that of SPEA-II, \textit{i.e.}, $\Gamma_A \preceq \Gamma_B$. Hence, for the rest of this study we focus on the non-dominated structures identified by NSGA-II.

%---------------------------------------------------------------------
\subsection{A posteriori Preference Articulation}
\label{s:resPrfA}

Each non-dominated structure essentially represents a varying degree of trade-off over search objectives, as seen in Fig.~\ref{f:PF}. Especially, the contradiction between the dynamic prediction error ($\mathcal{E}$) and the static error ($\overline{\mathcal{E}}$) is worth noting. It is clear that improvement in dynamic/static performance comes with a trade-off in the static/dynamic performance. This further highlights the need for a multi-objective approach.

For further analysis, 3 structures are selected from the identified non-dominated structures in $\Gamma_A$, following the \textit{a posteriori} selection approaches discussed in Section~\ref{s:PrA}. The selected structures and corresponding coefficients are given in the following models:

%---------------------------------------------------------------------
\begin{small}                                  
\begin{align}                               
\label{eq:M1}                               
& \mathcal{M}_1: y(k) =  12.047 + 0.9268 \ y(k-1) - 0.26037 \ y(k-3) \nonumber \\  
         & - 4.9214 \ u(k-2) + 1.0603 \ u^2(k-3) + 12.289 \ u^3(k-1) \nonumber\\
         & + 12.777 \ u^2(k-3)u(k-1) - 19.02 \ u(k-4)u(k-3)u(k-1) \nonumber\\ 
         & -12.831 \ u(k-3)u^2(k-2) + 13.662 \ u(k-4)u^2(k-2) \nonumber \\ 
         & + 5.366 \, u(k-4)u(k-3)u(k-2) - 6.1856 \ u^2(k-5)u(k-2) \nonumber \\ 
         & - 36.094 \ u(k-5)u^2(k-1) + 40.953 \ u^2(k-5)u(k-1) \nonumber \\
         & -11.064 \ u^3(k-5)\\
% \end{align}                                 
% \end{small} 
% %---------------------------------------------------------------------
% %---------------------------------------------------------------------
% \begin{small}                                  
% \begin{align}
\smallskip
\label{eq:M2}                                
& \mathcal{M}_2: y(k) = 21.366 + 0.76405 \ y(k-2) - 0.38755 \ y(k-4) \nonumber \\           
         & - 7.7188 \ u(k-2) - 4.086 \ u^2(k-1) + 2.5905 \, u(k-2)u(k-1) \nonumber \\ 
         & -2.2637 \, u(k-5)u^2(k-1) - 0.054858 \, u(k-5)u(k-4)u(k-1) \nonumber \\ 
         & + 2.8763 \ u^2(k-5) + 2.1183 \, u^3(k-1)\\
% \end{align}                                 
% \end{small}
% %--------------------------------------------------------------------- 
% %---------------------------------------------------------------------
% \begin{small}                                  
% \begin{align}                               
\label{eq:M3}                               
& \mathcal{M}_3: y(k) = 14.986 + 0.72049 \ y(k-1) - 0.12131 \, y(k-5) \nonumber \\           
         & -6.6797 \, u(k-2) + 1.6136 \, u^2(k-5) + 1.8557 \, u(k-2)u^2(k-1) \nonumber\\ 
         & -1.2517 \, u(k-5)u^2(k-1) -1.6357 \, u(k-3)u(k-2)u(k-1) \nonumber\\
         & + 0.80815 \, u^2(k-3)u(k-2)   
\end{align}                                 
\end{small}
%---------------------------------------------------------------------  
%%----------------------------------------------------------------------
\begin{table}[!t]
  \centering
  \small
  \caption{Selected Models}
  \label{t:ndsol}%
  \begin{adjustbox}{max width=0.45\textwidth}
  \begin{threeparttable}
    \begin{tabular}{ccccc}
    \toprule
    \textbf{Model} & \makecell{\textbf{Number of}\\ \textbf{Terms} (\boldmath$\xi$)} & \makecell{\textbf{Dynamic}\\ \textbf{Error} (\boldmath$\mathcal{E}$)}     & \makecell{\textbf{Static} \\ \textbf{Error} (\boldmath$\bar{\mathcal{E}}$)} & \textbf{Remark} \\
    \midrule
    $\mathcal{M}_1$ & 15  & 14.26 & 1.56 & \makecell{MMD,\\ $O_1$ + MTD}\\[1ex]
    \midrule
    $\mathcal{M}_2$ & 10  & 19.73 & 1.39 & $O_2$ + MTD \\[0.8ex]
    \midrule
    $\mathcal{M}_3$ & 9  & 16.80 & 2.39 & $O_3$ + MTD \\
    \bottomrule
    \end{tabular}%
  \end{threeparttable}
  \end{adjustbox}
\end{table}%
%%----------------------------------------------------------------------

The objective function values of the selected models are shown in Table~\ref{t:ndsol}. The first model $\mathcal{M}_1$ has been selected following the MMD approach (see Section~\ref{s:MMD}) and therefore represents the overall compromise. Further, three distinct scenarios for \textit{a posteriori} preference are considered to highlight the degree of compromise represented by non-dominated structures. In the first scenario, the dynamic and static performance are preferred over the cardinality (see $O_1$, Table~\ref{t:ObjRank}) which also leads to the selection of the model $\mathcal{M}_1$. Next, the parsimonious structure with a better static performance is preferred with a trade-off in the dynamic performance (see $O_2$, Table~\ref{t:ObjRank}). This leads to the selection of model $\mathcal{M}_2$. The parsimony is also preferred in the last scenario, albeit here  dynamic performance is assigned more weight in comparison to the static error (see $O_3$, Table~\ref{t:ObjRank}). The last model $\mathcal{M}_3$ encapsulates this scenario. 

The identified models are validated by the correlation based model-validity tests~\cite{Billings:2013}. The outcomes of these tests are shown in Table~\ref{t:corr} which shows that the identified models could satisfy all correlation tests.
%----------------------------------------------------------------------
\begin{table}[!t]
  \centering
  \small
  \caption{Correlation Based Model Validity Tests\cite{Billings:2013}}
  \label{t:corr}%
    \begin{tabular}{cccc}
    \toprule
    \textbf{Test} & \boldmath$\mathcal{M}_1$ & \boldmath$\mathcal{M}_2$ & \boldmath$\mathcal{M}_3$ \\[0.5ex]
    \midrule
    $\Phi_{\epsilon\epsilon}$   & \cmark      & \cmark      &  \cmark\\[0.2ex]
    $\Phi_{u\epsilon}$          & \cmark      & \cmark      &  \cmark\\[0.2ex]
    $\Phi_{u^{2}\epsilon}$      & \cmark      & \cmark      &  \cmark\\[0.2ex]
    $\Phi_{u^{2}\epsilon^2}$    & \cmark      & \cmark      &  \cmark\\[0.2ex]
    $\Phi_{\epsilon^{2}u}$      & \cmark      & \cmark      &  \cmark\\[0.2ex]
    \bottomrule
    \end{tabular}%
\end{table}%
%----------------------------------------------------------------------

%%-------------------------------------------------------------
\begin{table}[!t]
  \small
  \centering
  \caption{Coefficients of the Static Model}
  \label{t:staticcoeff}
    \begin{tabular}{ccccc}
    \toprule
    \multirow{2}[4]{*}{\textbf{Model}} & \multicolumn{4}{c}{\textbf{Coefficients}} \\
    \cmidrule{2-5}          & \boldmath{$a_0$} & \boldmath{$a_1$} & \boldmath{$a_2$} & \boldmath{$a_3$} \\ [0.5ex]
    \midrule
    $\mathcal{M}_1$    & 36.1141 & -14.7537 & 3.1786 & -0.4453 \\[0.5ex]
    $\mathcal{M}_2$    & 34.2686 & -12.3798 & 2.2145 & -0.3213 \\[0.5ex]
    $\mathcal{M}_3$    & 37.3892 & -16.6653 & 4.0258 & -0.5578 \\[0.5ex]
    \bottomrule
    \end{tabular}%
\end{table}%
%%-------------------------------------------------------------
%%-------------------------------------------------------------
\subsection{Steady State Relation of The Identified Models}
\label{s:SS}

Given that in this study degree of nonlinearity ($n_l$) is fixed to $3$, the static input-output relation given in~(\ref{eq:clusterPoly}) can further be simplified as follows:
%-------------------------------------------------------------
\begin{align}
\label{eq:cluster5}
     \overline{y} = & a_0 + a_1 \overline{u} + a_2 \overline{u}^2 + + a_3 \overline{u}^3, \quad \text{where,}\\
     a_0= & \frac{\Sigma_0}{1-\Sigma_y}, a_1=\frac{\Sigma_u}{1-\Sigma_y}, a_2=\frac{\Sigma_{u^2}}{1-\Sigma_y}, a_3=\frac{\Sigma_{u^3}}{1-\Sigma_y} \nonumber
\end{align}
%-------------------------------------------------------------
This gives general form of steady state relation of the identified models. The coefficients of~(\ref{eq:cluster5}) are dependent both on terms and the corresponding coefficients of the identified models. For the selected models, $\mathcal{M}_1$-$\mathcal{M}_3$, these coefficients are shown in Table~\ref{t:staticcoeff}. 

%%-------------------------------------------------------------
\begin{table}[!t]
  \small
  \centering
  \caption{Coefficients Estimated by OFR~\cite{Aguirre:Donoso:2000} and OFR-EA~\cite{Correa:Aguirre:2002}}
  \label{t:luiscoeff}
    \begin{tabular}{ccc}
    \toprule
    \textbf{Coefficients} & \textbf{OFR} & \textbf{OFR-EA} \\
    \midrule
    $\theta_0$ & 6.2479 & 13.7292 \\[0.5ex]
    $\theta_1$ & 1.2013 & 0.7315 \\[0.5ex]
    $\theta_2$ & -0.2608 & -0.0047 \\[0.5ex]
    $\theta_3$ & -2.6783 & -0.8280 \\[0.5ex]
    $\theta_4$ & -0.2080 & -0.2495 \\[0.5ex]
    $\theta_5$ & 8.8399 & 3.6774 \\[0.5ex]
    $\theta_6$ & 3.6636 & 2.0210 \\[0.5ex]
    $\theta_7$ & -0.6162 & -1.7617 \\[0.5ex]
    $\theta_8$ & -9.7707 & -4.6409 \\[0.5ex]
    \bottomrule
    \end{tabular}%
\end{table}%
%%-------------------------------------------------------------
%---------------------------------------------------------------------
\subsection{Comparative Evaluation}
\label{s:res3}

For the purpose of the comparative evaluation, the selected models (\textit{i.e.}, $\mathcal{M}_1$, $\mathcal{M}_2$ and $\mathcal{M}_3$) are compared with the models identified for the same experimental setup and the identification data of the buck converter by a different grey-box identification approaches: OFR~\cite{Aguirre:Donoso:2000} and OFR-EA~\cite{Correa:Aguirre:2002}. The models identified in these earlier investigations are as follows:
%---------------------------------------------------------------------
% \begin{small}
\begin{align}
    \label{eq:luismodel}
    y(k) & = \theta_0 + \theta_1 y(k-1) + \theta_2 y(k-2) + \theta_3 u^3(k-1)\nonumber\\
         & + \theta_4 y(k-3) + \theta_5 u^2(k-1)u(k-3) + \theta_6 u^3(k-3)\nonumber\\
         & + \theta_7 u(k-1)u(k-3) + \theta_8 u^2(k-3)u(k-1)
\end{align}
% \end{small}
%---------------------------------------------------------------------
The corresponding coefficients `$\theta$' are given in Table~\ref{t:luiscoeff}.

\begin{figure}[!t]
\centering
\small
\begin{subfigure}{.42\textwidth}
  \centering
  \includegraphics[width=\textwidth]{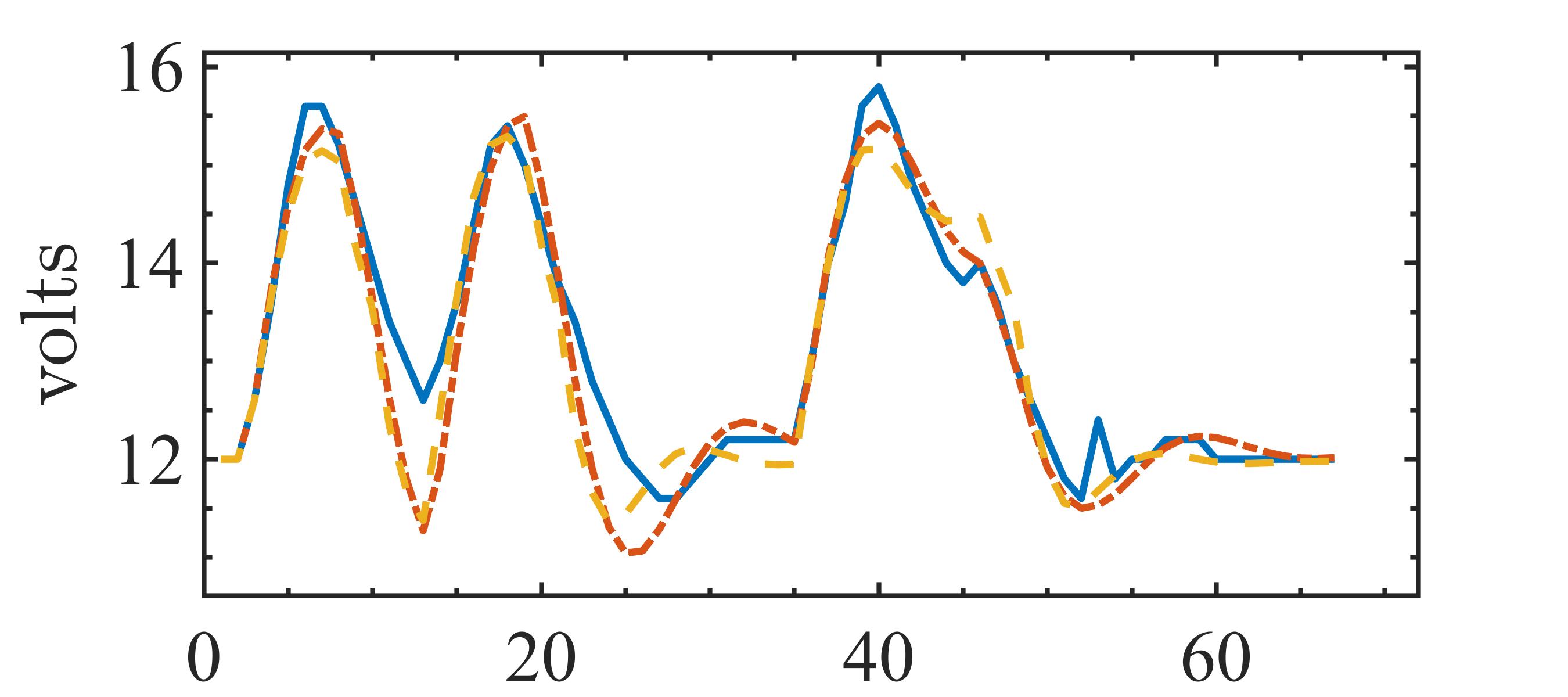}
  \caption{$(-)$ Validation data; $(-\cdotp)$ OFR\cite{Aguirre:Donoso:2000}; $(--)$ OFR-EA\cite{Correa:Aguirre:2002}.}
  \label{f:ympo_ofr}
\end{subfigure}%
\hfill
% \hspace{0.1\textwidth}
\begin{subfigure}{.42\textwidth}
  \centering
  \includegraphics[width=\textwidth]{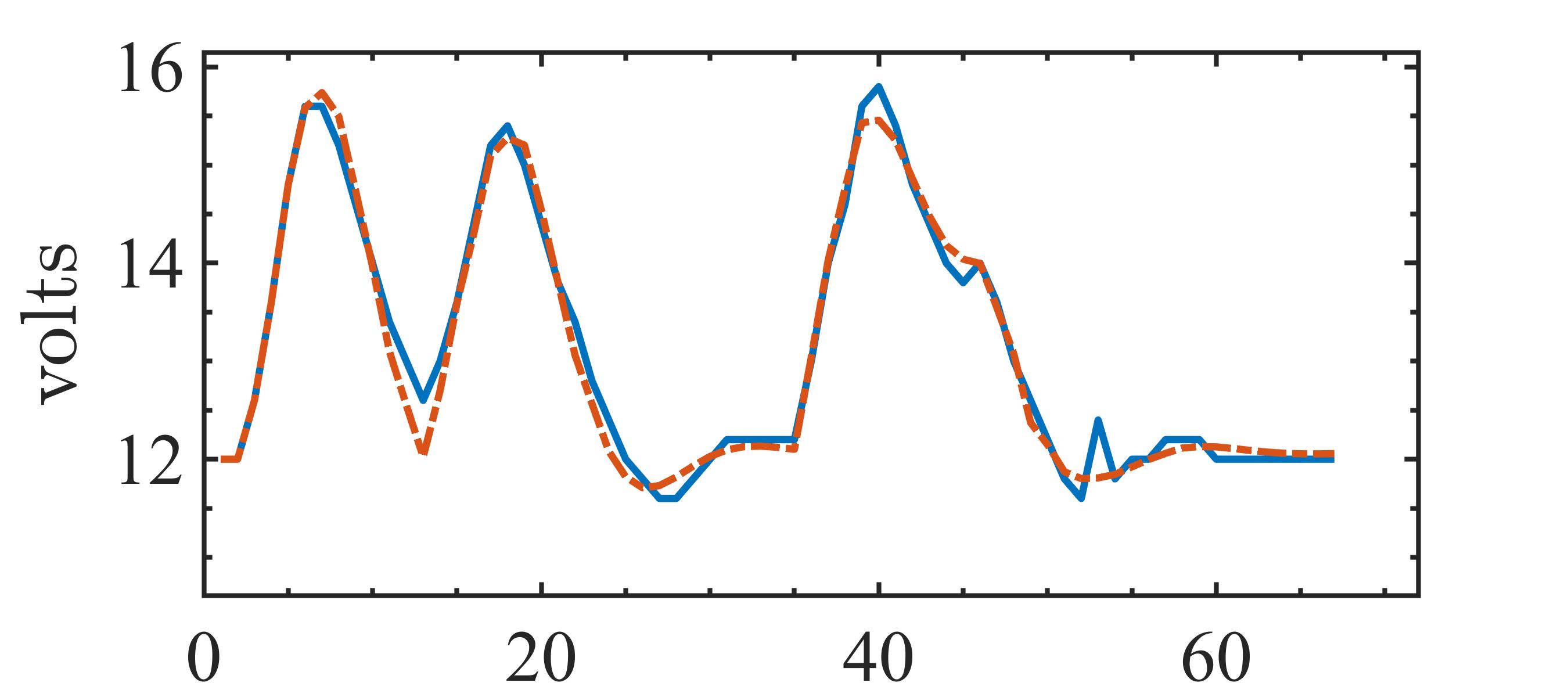}
  \caption{$(-)$ Validation data; $(-\cdotp)$ $\mathcal{M}_1$.}
  \label{f:ympo_model1}
\end{subfigure}

\begin{subfigure}{.42\textwidth}
  \centering
  \includegraphics[width=\textwidth]{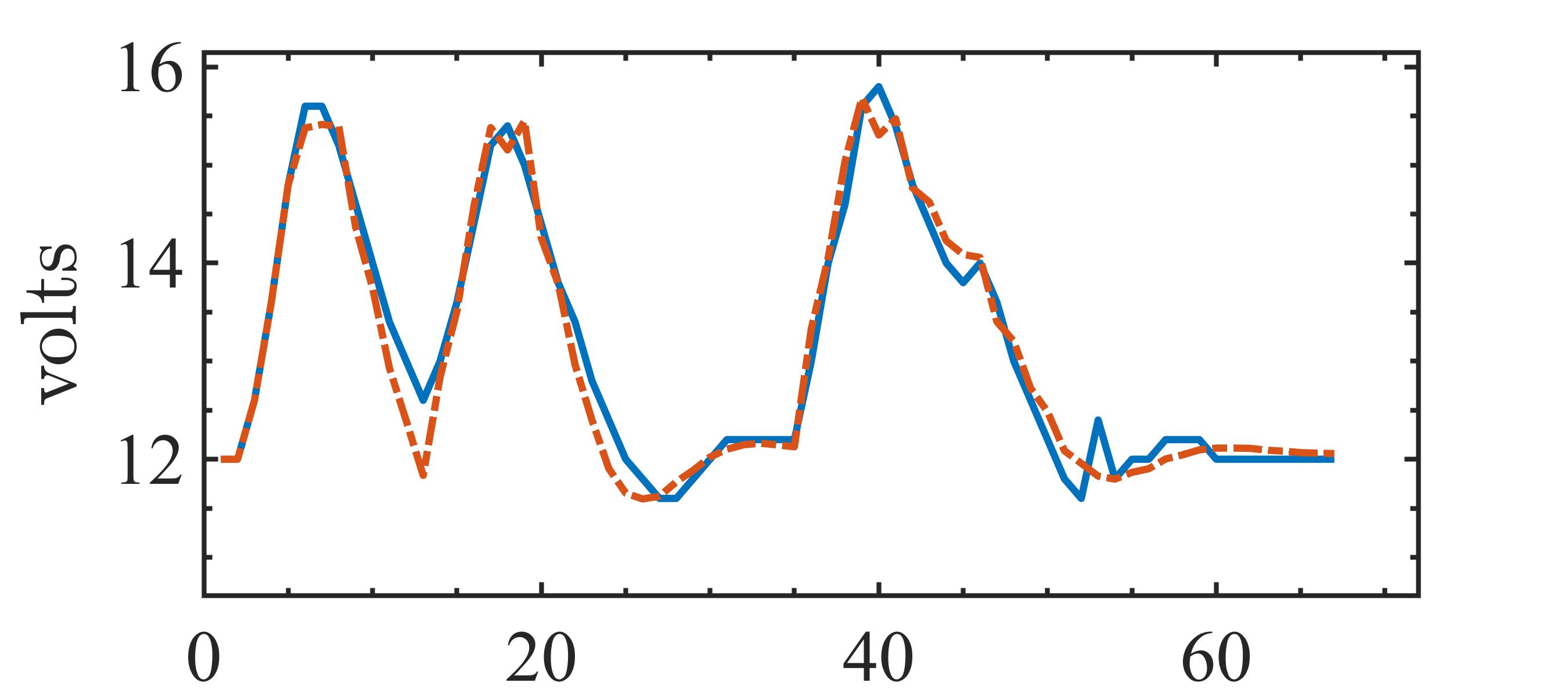}
  \caption{$(-)$ Validation data; $(-\cdotp)$ $\mathcal{M}_2$.}
  \label{f:ympo_model2}
\end{subfigure}%
% \hfill
\hspace{0.1\textwidth}
\begin{subfigure}{.42\textwidth}
  \centering
  \includegraphics[width=\textwidth]{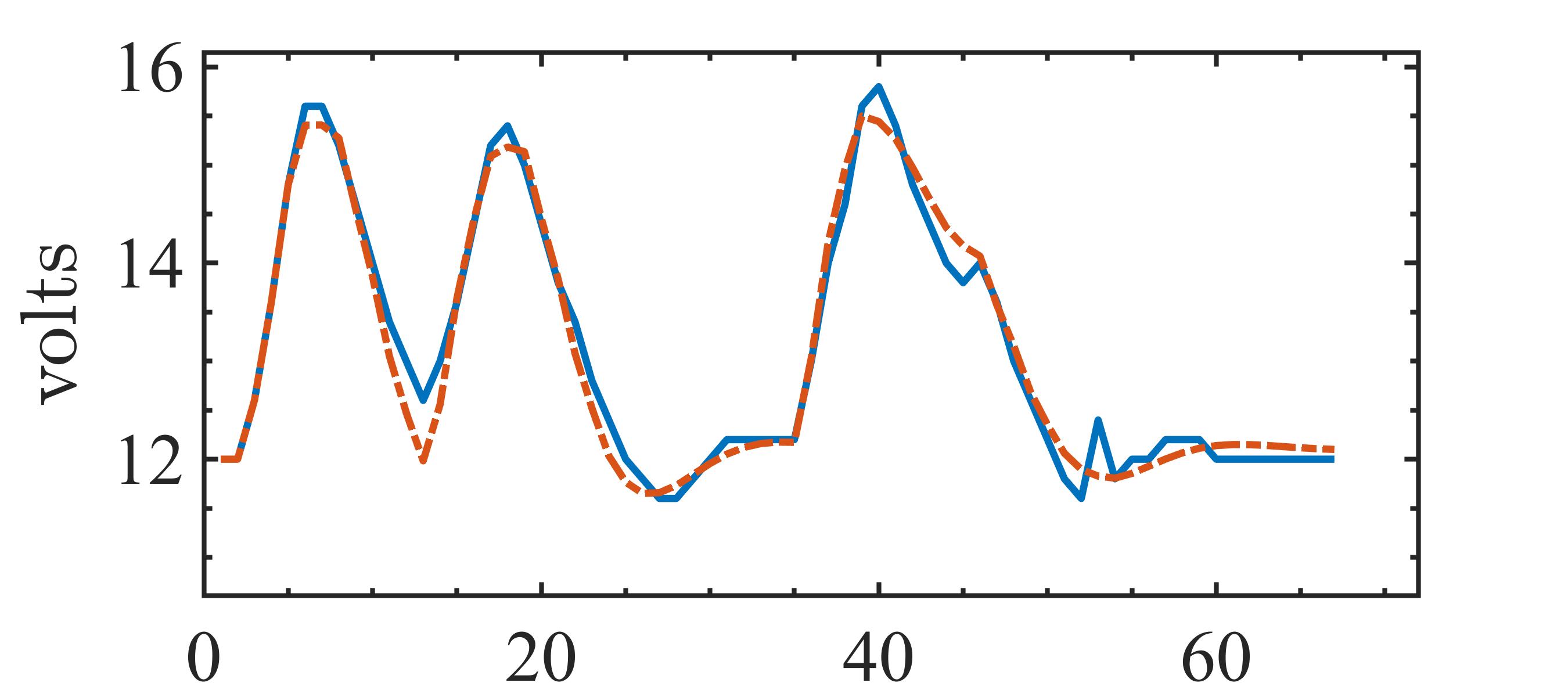}
  \caption{$(-)$ Validation data; $(-\cdotp)$ $\mathcal{M}_3$.}
  \label{f:ympo_model3}
\end{subfigure}
\caption{Model predicted output ($\hat{y}$) over validation data. The dynamic prediction error ($\mathcal{E}$) of the models is as follows: $35.01\%$ (OFR);  $33.13\%$ (OFR-EA); $14.26\%$ ($\mathcal{M}_1$); $19.73\%$ ($\mathcal{M}_2$); $16.80\%$ ($\mathcal{M}_3$)}
\label{f:ympo}
\end{figure}

First, the dynamic prediction capability of the models is compared by calculating the model-predicted output over the validation data, as shown in Fig.~\ref{f:ympo}. It is clear that the models identified using the proposed approach could yield comparatively better prediction performance. The prediction error with the identified models lie in the range of $[14\%-20\%]$. In comparison, OFR~\cite{Aguirre:Donoso:2000} and OFR-EA~\cite{Correa:Aguirre:2002} could yield approximately $33\%$ prediction error; clearly a higher trade-off is made in the dynamic performance with these approaches.

\begin{figure}[!t]
\centering
\small
\begin{subfigure}{.241\textwidth}
  \centering
  \includegraphics[width=\textwidth]{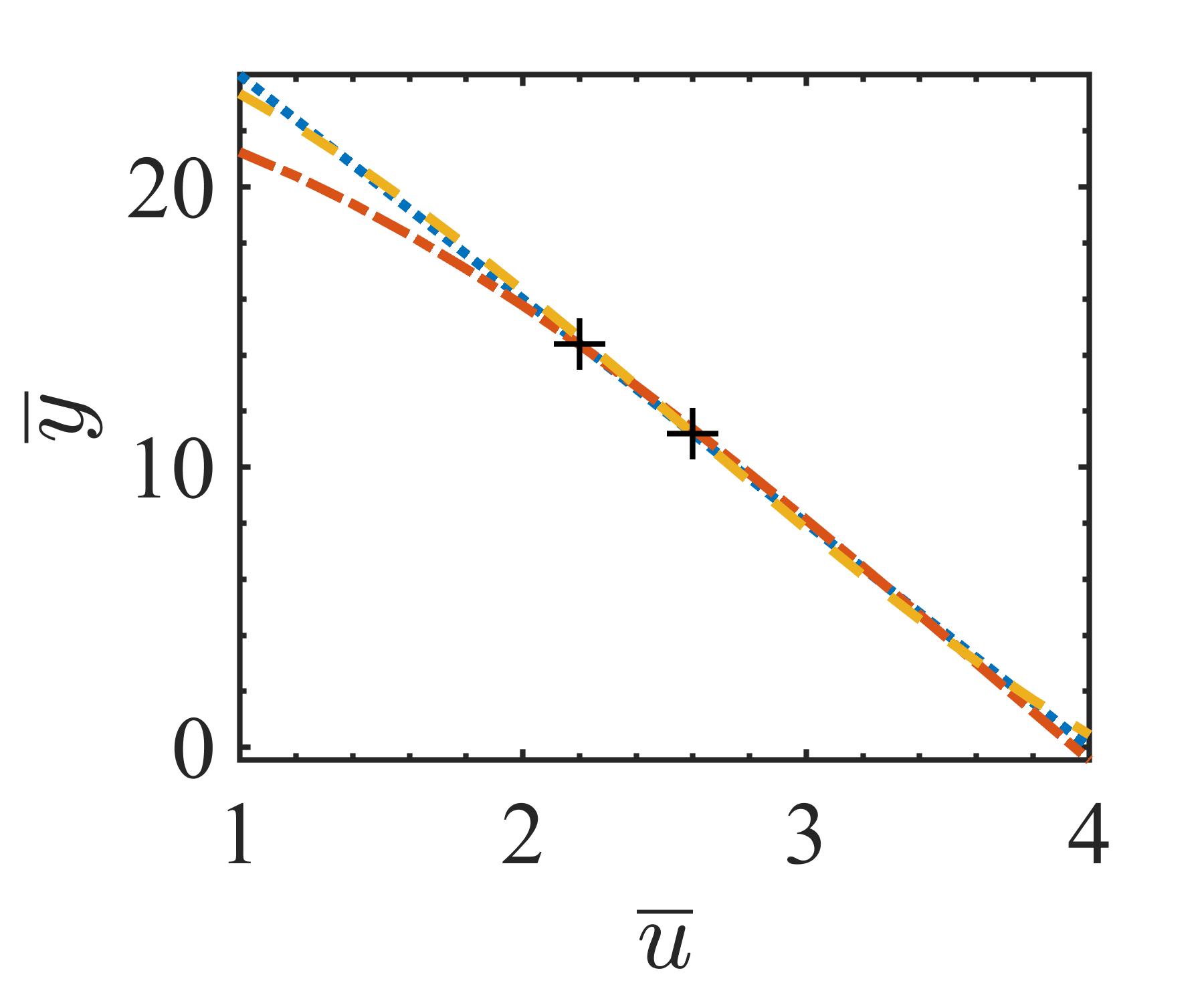}
  \caption{$(\cdotp)$ Buck converter; $(-\cdotp)$ OFR\cite{Aguirre:Donoso:2000}; $(-)$ OFR-EA\cite{Correa:Aguirre:2002}.}
  \label{f:ys1}
\end{subfigure}%
\hfill
% \hspace{0.15\textwidth}
\begin{subfigure}{.241\textwidth}
  \centering
  \includegraphics[width=\textwidth]{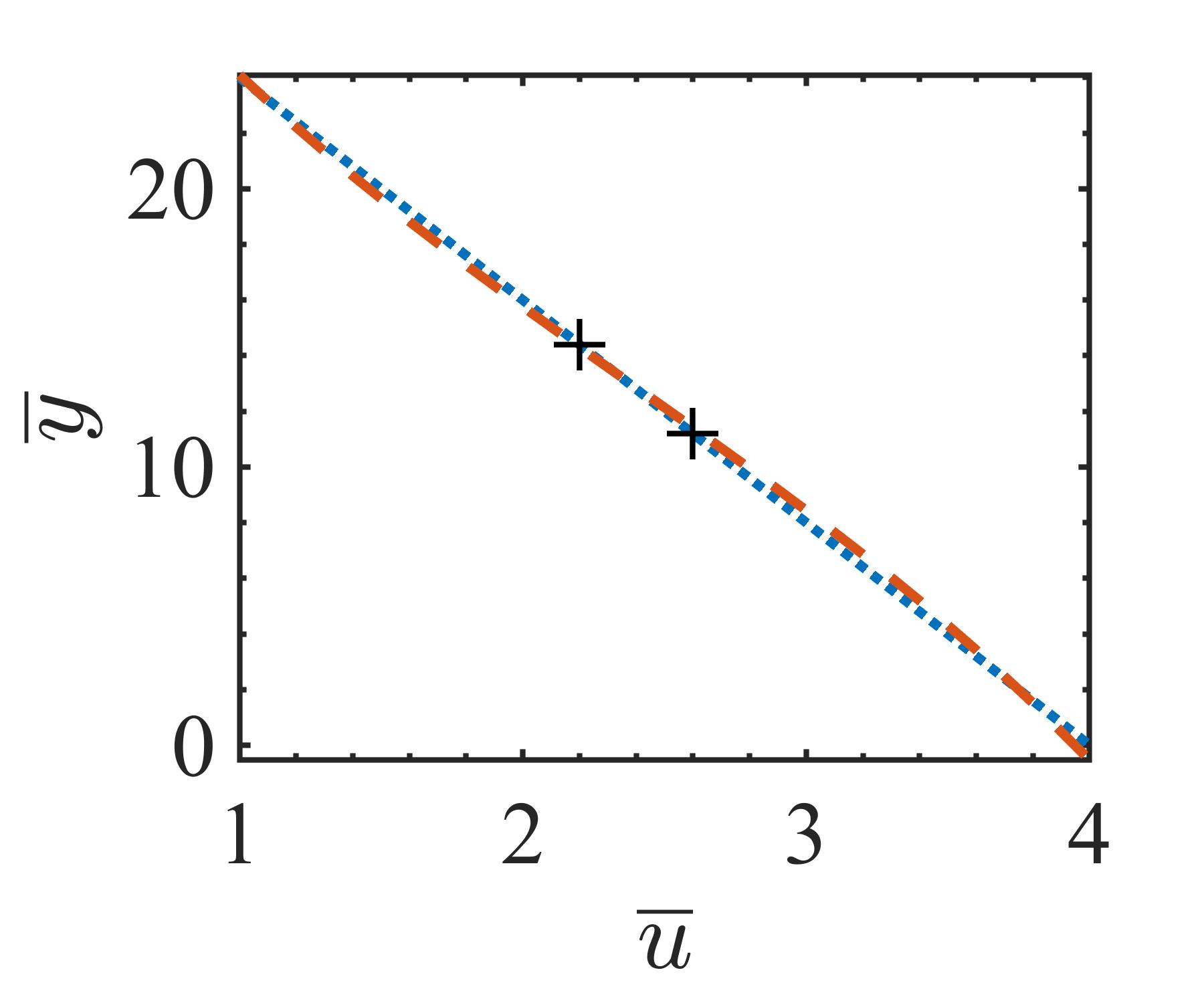}
  \caption{$(\cdotp)$ Buck converter; $(-)$ $\mathcal{M}_1$.}
  \label{f:ys2}
\end{subfigure} 
\\
\begin{subfigure}{.241\textwidth}
  \centering
  \includegraphics[width=\textwidth]{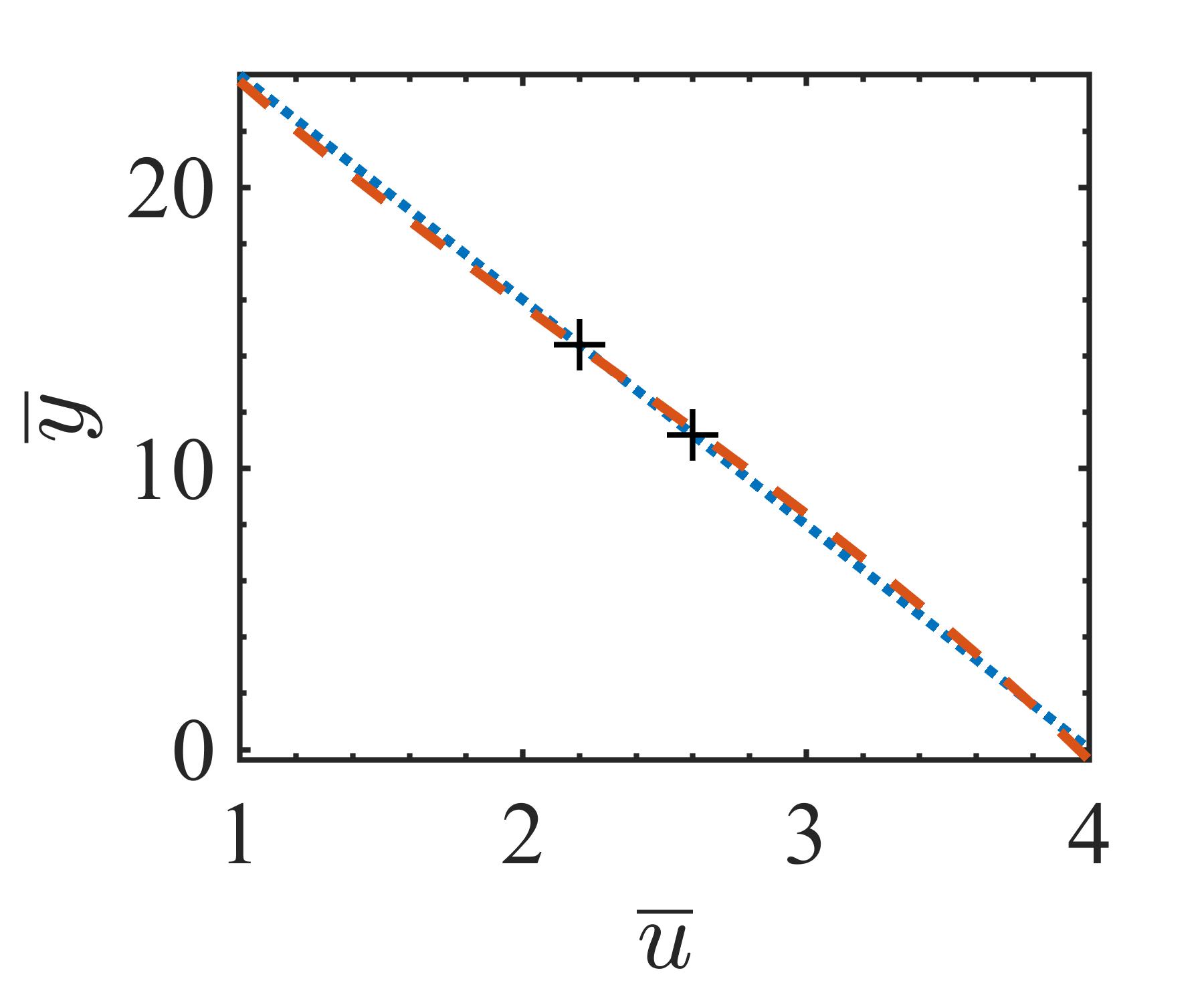}
  \caption{$(\cdotp)$ Buck converter; $(-)$ $\mathcal{M}_2$}
  \label{f:ys3}
\end{subfigure}
\hfill
% \hspace{0.15\textwidth}
\begin{subfigure}{.241\textwidth}
  \centering
  \includegraphics[width=\textwidth]{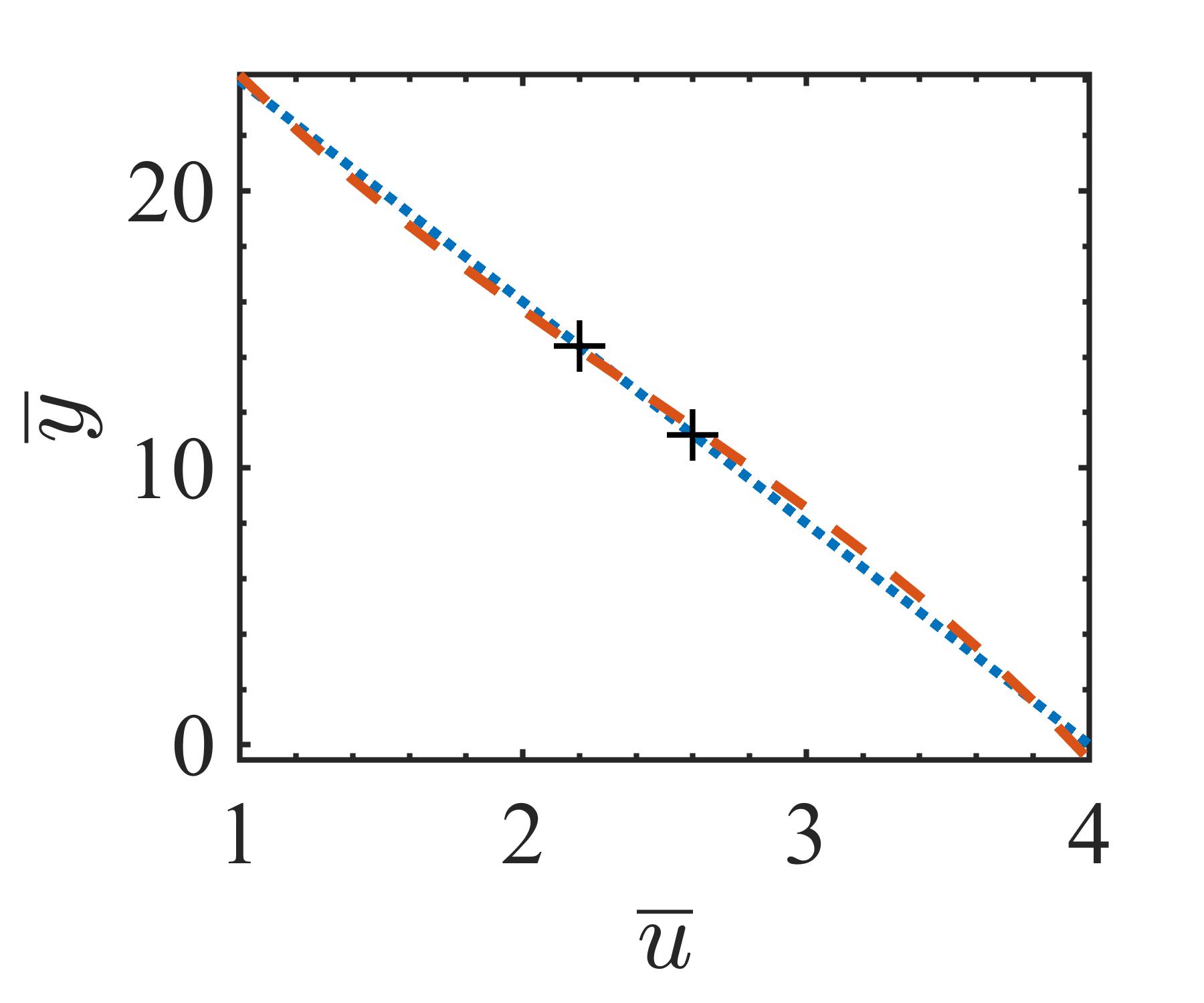}
  \caption{$(\cdotp)$ Buck converter; $(-)$ $\mathcal{M}_3$.}
  \label{f:ys4}
\end{subfigure}%
\caption{Static function of the buck converter and the identified models. `$+$' sign indicates the range of the input contained in the identification data, \textit{i.e.}, $[2.2V-2.5V]$. The error obtained in the static behavior ($\overline{\mathcal{E}}$) is as follows:$15.21$ (OFR); $1.21$ (OFR-EA); $1.56$ ($\mathcal{M}_1$); $1.39$ ($\mathcal{M}_2$); $2.39$ ($\mathcal{M}_3$)}
\label{f:ystatic}
\end{figure}

Next, the static behavior of the models is evaluated as shown in Fig.~\ref{f:ystatic}. It is worth noting that, while the static behavior is evaluated over the valid input range of $[1V-4V]$, the identification data has been generated over the relatively narrow input range of $2.2V \leq u(k) \leq 2.5 V$. Therefore, evaluation of the models beyond this input range can be considered as the evaluation of \textit{global} validity. As seen in Fig.~\ref{f:ys2}-\ref{f:ys4}, the identified models mimic the static behavior of the buck converter over almost the entire valid input range. Further, the identified models yield the static error $\bar{\mathcal{E}}$ in the range of $[1.39-2.39]$, which is better than/comparable to OFR/OFR-EA.

Further, the degree of compromise over the search objectives is clearly visible in the dynamic and static behavior of the identified models. For example, among the identified models, the prediction capability of $\mathcal{M}_2$ is comparatively poor with $\mathcal{E}=19.7\%$, as seen in Fig.~\ref{f:ympo_model2}. However, with this trade-off, $\mathcal{M}_2$ could perfectly mimic the static behavior of the buck converter over the entire input range, as seen in Fig.~\ref{f:ys3}. 

Nevertheless, it is interesting to see that all the identified models ($\mathcal{M}_1-\mathcal{M}_3$) yield practically acceptable dynamic and static performance. Therefore, the selection of final model from $\mathcal{M}_1-\mathcal{M}_3$ is subjective and dependent on the DM's preference. To this end, without the loss of generality, the principle of parsimony is followed in this study. Since $\mathcal{M}_3$ provides relatively compact description of the system dynamics, it is recommended to model the buck converter.
%%----------------------------------------------------------------------
%----------------------------------------------------------------------
%%----------------------------------------------------------------------
%----------------------------------------------------------------------

%---------------------------------------------------------------------
\subsection{Role of Non-linear Input Clusters: Some Comments}
\label{s:CommTC}

In this study, prior to the structure selection, the \textit{nonlinear output} and \textit{cross-term} clusters are removed from $\mathcal{X}_{model}$, as discussed earlier in Section~\ref{s:useofprior}. Further, a closer inspection of the static input-output relations in~(\ref{eq:static2}) and~(\ref{eq:clusterPoly}) shows that only the following three term clusters are required to induce the `\textit{perfect}' static behavior of buck converter: \textit{constant terms} ($\Omega_0$), \textit{linear input} ($\Omega_u$) and \textit{linear output} ($\Omega_y$). Thus, if the terms belonging to the \textit{nonlinear-input clusters} (\textit{i.e.}, $\Omega_u^p, p=2,\dots n_l$) are also removed, then~(\ref{eq:clusterPoly}) simplifies to,

%-------------------------------------------------------------
\begin{align}
\label{eq:clusterPoly2}
     \overline{y} = & a_0 + a_1 \overline{u}\\
     \text{where, } a_0= & \frac{\Sigma_0}{1-\Sigma_y}, a_1=\frac{\Sigma_u}{1-\Sigma_y}\nonumber
\end{align}
%-------------------------------------------------------------
It is clear that this simplified static relation is similar to the static behavior of the buck converter in~(\ref{eq:static2}). 
%It is therefore apparently obvious that the \textit{non-linear input} clusters are not required to encode the static behavior. 
This could also be explained by the `\textit{straight-line}' nature of the static input-output relationship.

However, it is interesting to see that all the identified models, $\mathcal{M}_1-\mathcal{M}_3$, contain the terms from the \textit{non-linear input} clusters ($\Omega_u^p$). This implies that while the $\Omega_u^p$ clusters are not required for the static behavior, they may be essential for the dynamic prediction. 

To further investigate the role of $\Omega_u^p$ clusters, consider the identification of buck converter with the similar procedure, outlined in Algorithm~\ref{al:moss}, except with one key difference: In these experiments the non-linear input clusters are also removed, \textit{i.e.}, $\mathcal{X}_{model}=\{ \Omega_0 \cup \Omega_u \cup \Omega_y \}$. The model identified following this procedure is as follows:
%-------------------------------------------------------------------
% \begin{small}
\begin{align}
    \label{eq:linear}
    \mathcal{M}_{4}: y(k)= & \ 30.392 + 0.061677 \ y(k-3) \nonumber\\ 
          & - 5.6359 \ u(k-2) \ -1.8699 \ u(k-3) \nonumber\\ 
          & - 0.080413 \ u(k-4)  
\end{align}
% \end{small}
%-------------------------------------------------------------------

\begin{figure}[!t]
\centering
\small
\begin{subfigure}{.24\textwidth}
  \centering
  \includegraphics[width=\textwidth]{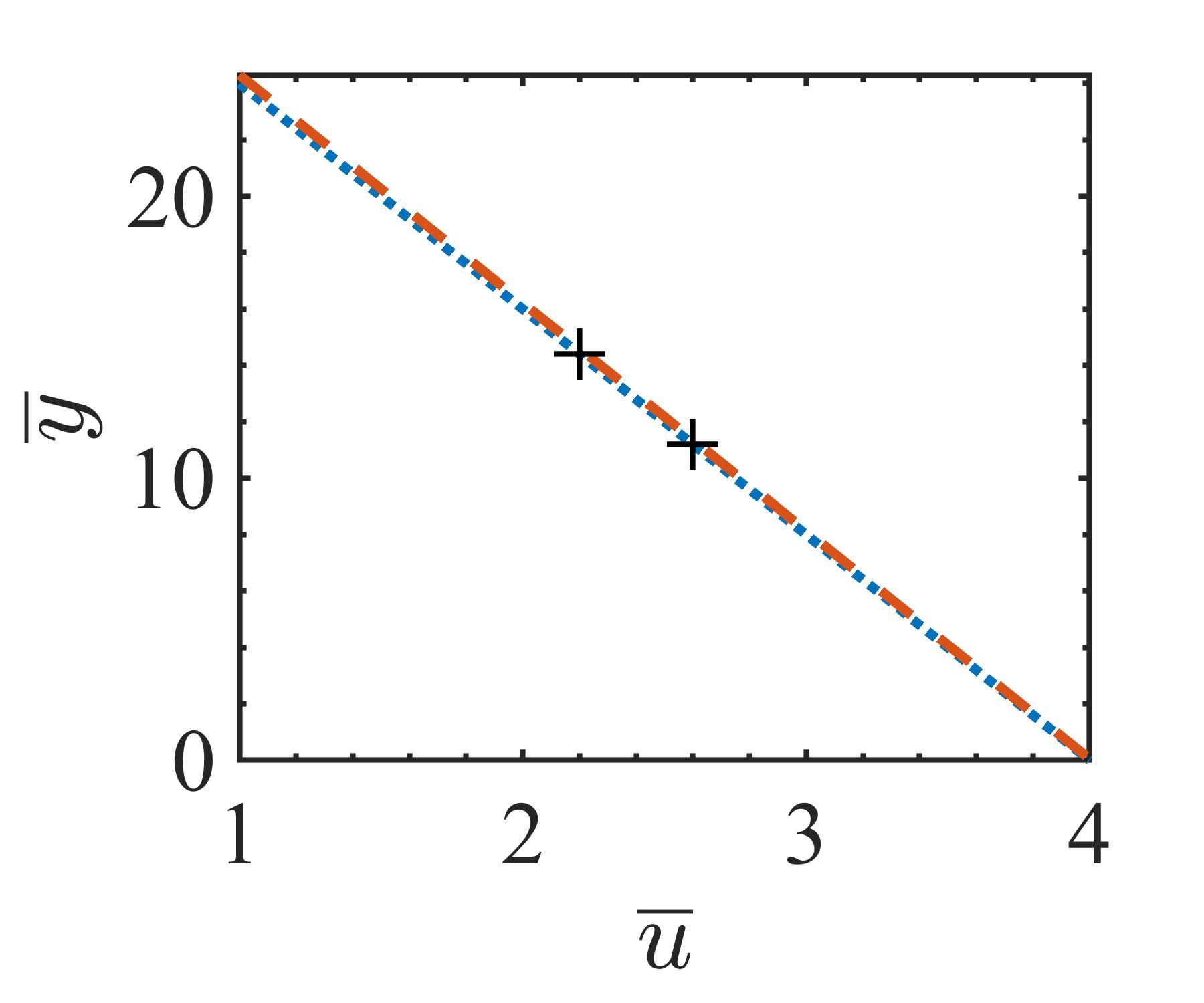}
  \caption{$(-)$ Validation data; $(-\cdotp)$ $\mathcal{M}_4$.}
  \label{f:static_linear}
\end{subfigure}
\hfill
% \hspace{0.1\textwidth}
\begin{subfigure}{.4\textwidth}
  \centering
  \includegraphics[width=\textwidth]{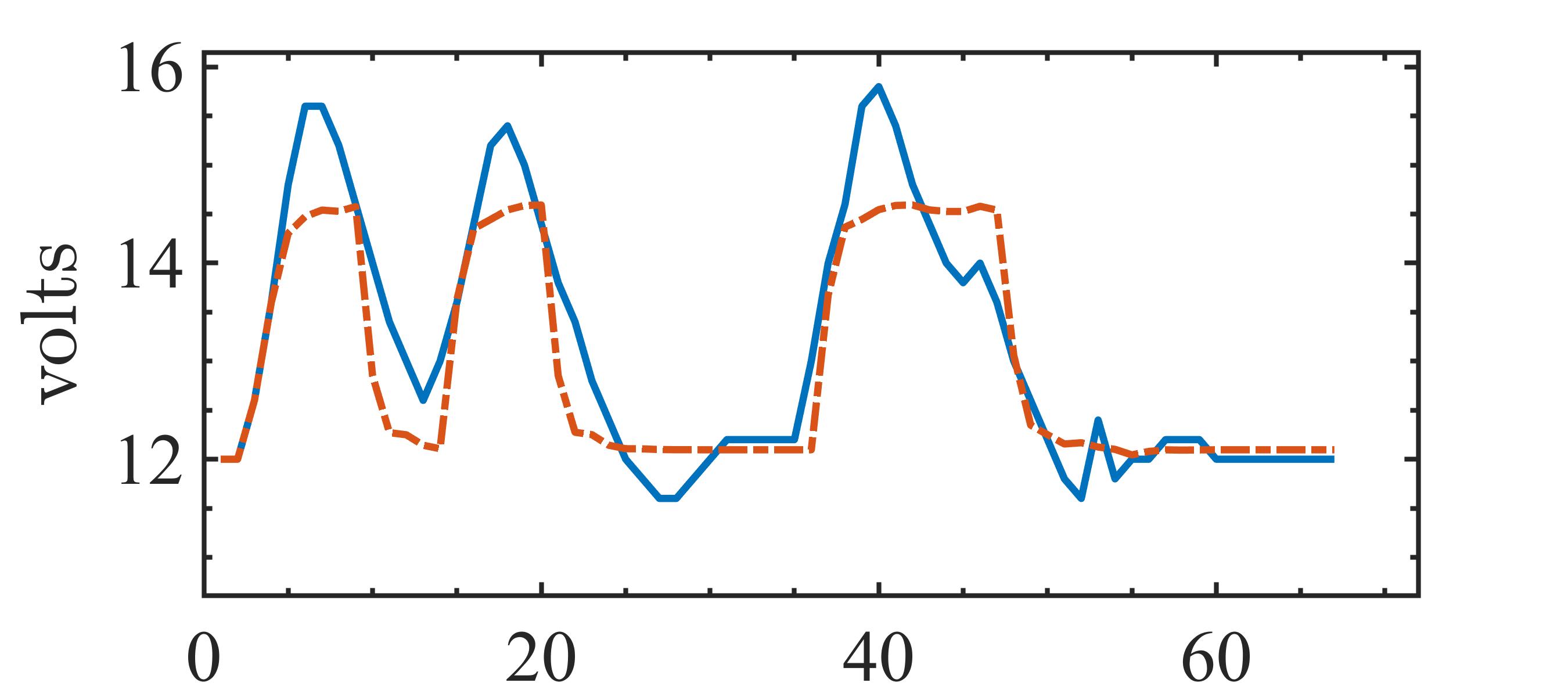}
  \caption{$(-)$ Validation data; $(-\cdotp)$ $\mathcal{M}_4$.}
  \label{f:ympo_linear}
\end{subfigure}%

\caption{The validation results obtained without non-linear input cluster, $\mathcal{M}_4$. (a) Static function of the buck converter and the linear model. `$+$' sign indicates the range of the input contained in the identification data, \textit{i.e.}, $[2.2V-2.5V]$. The error obtained in the static behavior ($\overline{\mathcal{E}}$) is $0.59$. (b) Model predicted output ($\hat{y}$) over validation data. The dynamic prediction error ($\mathcal{E}$) is $42.26\%$.}
\label{f:linear}
\end{figure}

The validation results for $\mathcal{M}_{4}$ are shown in Fig.~\ref{f:linear}. As expected, this model mimics the static behavior of buck converter very well, as seen in Fig.~\ref{f:static_linear}. This improvement, however, comes with a significant trade-off in the dynamic prediction capabilities, as seen in Fig.~\ref{f:ympo_linear}. This empirical results, therefore, confirms that it is necessary to include \textit{non-linear input} clusters ($\Omega_u^p$) into the model to improve the dynamic prediction.

%Further, in comparison to $\mathcal{M}_1-\mathcal{M}_3$, this model yields better static performance with $\overline{\mathcal{E}}=0.59$. %%------------------------------------------------------------------
%-------------------------------------------------------------------
%---------------------------------------------------------------------
\section{Conclusion}
\label{s:con}

A new multi-objective structure selection approach has been proposed which explicitly quantifies and uses the \textit{a priori} knowledge into the search process. The identification of buck converter dynamics is considered as a case study. A two-pronged approach is taken to embed \textit{a priori} information about the known static nonlinearity of the buck converter: 1) Set of candidate NARX terms is restricted. 2) Static behavior of the candidate structures is quantified and explicitly used as one of the search objectives. The results of this study convincingly demonstrate that the proposed approach can effectively utilize \textit{a priori} knowledge to identify parsimonious models with accurate dynamic prediction capabilities while preserving the steady-state characteristic of the system over a wide input range. 

\appendices
% \small
\section{Illustrative Example: Solution Representation}
\label{s:aSr}
Consider a simple NARX model with a total of $5$ terms ($n=5$) as follows:
%-------------------------------------------------------
\begin{align}
\label{eq:NARXExample}
	\mathcal{X}_{model} & = \begin{bmatrix} x_1 & x_2 & x_3 & x_4 & x_5 \end{bmatrix} \\
			  & =\begin{bmatrix} y(k-1) & u(k-1) & y(k-2)^2 \cdots \\
			                      & \cdots y(k-2)u(k-2) & u(k-3)^3 \end{bmatrix} \nonumber	  
\end{align}
%-------------------------------------------------------
For this problem, assume that the position of the $i^{th}$ particle is given by,
%---------------------------------------------------------
\begin{align}
\label{eq:NARXExample1}
    \beta_i & = \begin{bmatrix} 1 & 0 & 0 & 1 & 1 \end{bmatrix}
\end{align}
%---------------------------------------------------------
This implies that only the \textit{first, fourth} and \textit{fifth} terms from the set $\mathcal{X}_{model}$ are included into the \textit{structure/term subset}. Thus, the structure `$\mathcal{X}_i$' encoded by the particle $\beta_i$ is given by,
%---------------------------------------------------------
\begin{align*}
% \label{eq:NARXExample2}
    \mathcal{X}_i & = \begin{bmatrix} x_1 & x_4 & x_5 \end{bmatrix} \\
                  & = \begin{bmatrix} y(k-1) & y(k-2)u(k-2) & u(k-3)^3\end{bmatrix}
\end{align*}
%---------------------------------------------------------
%------------------------------------------------------------------
\section{Illustrative Example: Priority Weights}
\label{s:aPrw}

Let the objective rankings and the preference intensity specified by the DM be given by: $[O_\xi, O_\mathcal{E}, O_{\overline{\mathcal{E}}}] = [3,1,2]$ and $\mathcal{I}=5$. The corresponding multiplicative preference relations can be determined as follows (\textit{see} Line~\ref{line:mtd1}-\ref{line:mtd2}, Algorithm~\ref{al:mtd}):
%------------------------------------------------------------------
\begin{align*}
    % \label{eq:prefExample1}
    \begin{bmatrix} 
        \tau_{\xi,\xi} & \tau_{\xi,\mathcal{E}} & \tau_{\xi,\overline{\mathcal{E}}}\\ \tau_{\mathcal{E},\xi} & \tau_{\mathcal{E},\mathcal{E}} & \tau_{\mathcal{E},\overline{\mathcal{E}}}\\
        \tau_{\overline{\mathcal{E}},\xi} & \tau_{\overline{\mathcal{E}},\mathcal{E}} & \tau_{ \overline{\mathcal{E}},\overline{\mathcal{E}}}
        \end{bmatrix} = \begin{bmatrix} 
                        1 & \frac{1}{5} & \frac{1}{\sqrt{5}} \\
                        5 & 1 & \sqrt{5} \\ 
                        \sqrt{5} & \frac{1}{\sqrt{5}} & 1 
                        \end{bmatrix}%\nonumber
\end{align*}
%------------------------------------------------------------------
Consequently, the preference weights are determined as follows (\textit{see} Line~\ref{line:mtd3}, Algorithm~\ref{al:mtd}):
%------------------------------------------------------------------
\begin{align*}
    % \label{eq:prefExample2}
    \begin{bmatrix} w_\xi \\ w_\mathcal{E} \\ w_{\overline{\mathcal{E}}}\end{bmatrix} & =\begin{bmatrix}  0.4472 \\ 2.2361 \\ 1 \end{bmatrix}\\%\nonumber
    \text{which yields, } \vec{w} & = \frac{\begin{bmatrix} w_\xi & w_\mathcal{E} & w_{\overline{\mathcal{E}}}\end{bmatrix}^T}{\sum w} = \begin{bmatrix} 0.1214 \\ 0.6071 \\ 0.2715\end{bmatrix} \\
    % & = \begin{bmatrix} 0.1214 & 0.6071 & 0.2715\end{bmatrix}^T 
\end{align*}
%------------------------------------------------------------------

% \section*{Acknowledgment}

\ifCLASSOPTIONcaptionsoff
  \newpage
\fi



%
\bibliographystyle{IEEEtran}
% \bibliography{ref}


\section*{Authors' Biographies}
\begin{wrapfigure}{l}{25mm} 
\includegraphics[width=1in,height=1.25in,clip,keepaspectratio]{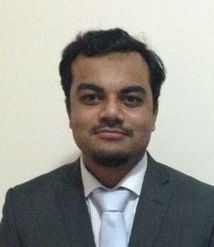}
\end{wrapfigure} \par
\textbf{Faizal Hafiz} received M.Tech degree in electrical engineering from National Institute of Technology (SVNIT), Surat, India in 2008.  From 2008 to 2010, he was working as an Assistant Manager at Reliance Infrastructure Ltd., Mumbai, India. From 2010 to 2016, he was a research Assistant Professor at King Saud University, Riyadh. He is currently a New Zealand International Doctoral Scholar and pursuing a doctoral degree at The University of Auckland, New Zealand in the field of Computational Intelligence and Control. His research interests include swarm intelligence, meta-heuristics and their applications to nonlinear control and signal processing.\par
\vspace{1.2mm}

\begin{wrapfigure}{l}{25mm} \includegraphics[width=1in,height=1.25in,clip,keepaspectratio]{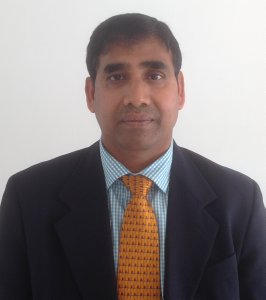}
\end{wrapfigure}\par

\textbf{Akshya Swain}~(M’97–SM’13) received Bsc(Eng) \& ME degree in 1985 and 1988 and Ph.D. degree in Control Engineering from The University of Sheffield in 1996. He has authored over 200 papers in International journals and conferences. Dr Swain is an Associate Editor of IEEE Sensors Journal and Member of the Editorial Board of International Journal of Automation and Control, International Journal of Sensors and Wireless Communications and Control. His research interests include nonlinear system identification and control, machine learning \& big data.\par

\vspace{1.2mm}

\begin{wrapfigure}{l}{25mm} 
\includegraphics[width=1in,height=1.25in,clip,keepaspectratio]{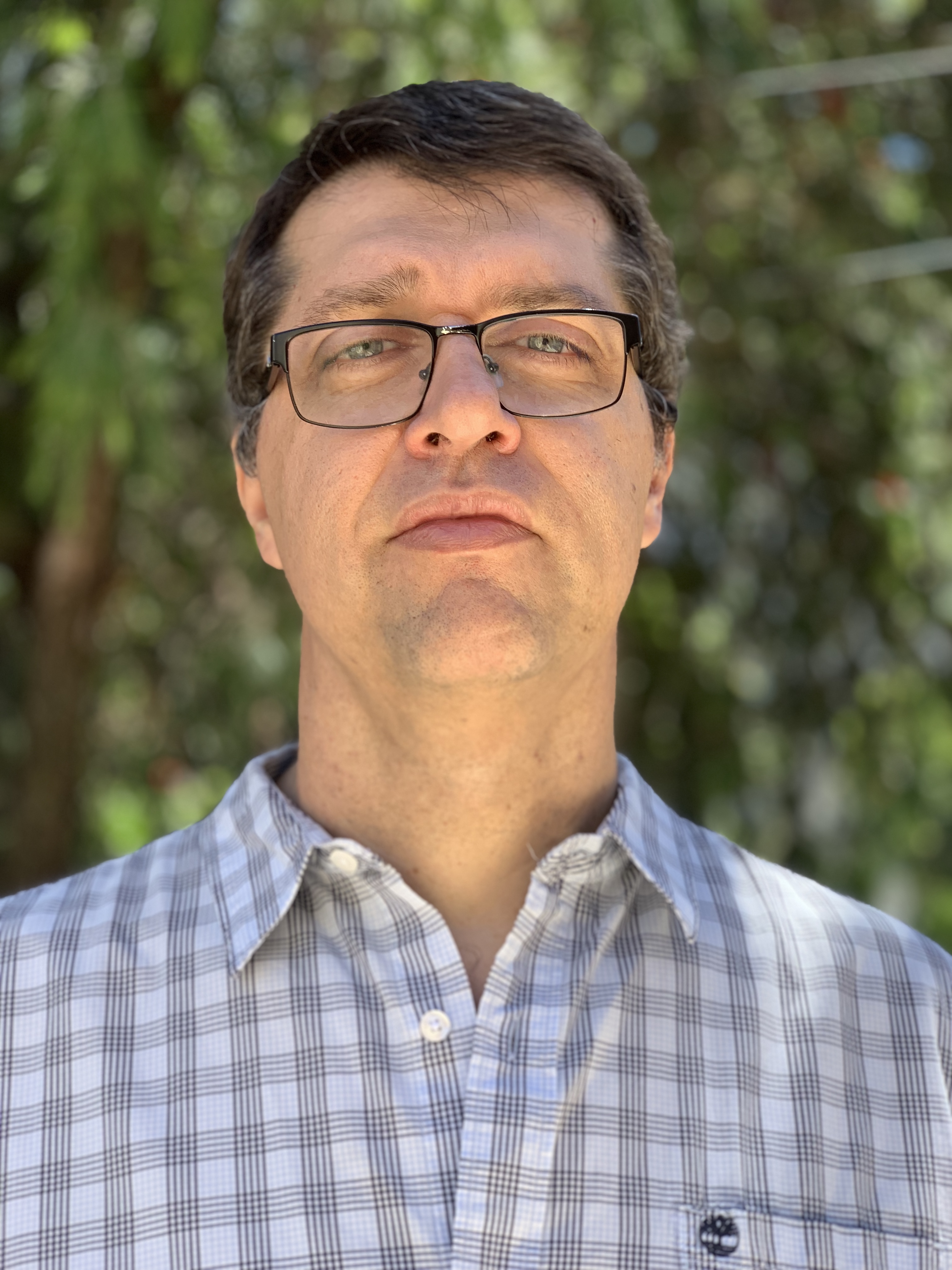}
\end{wrapfigure}\par
\textbf{Eduardo M. A. M. Mendes} received the B.Eng. degree in electrical engineering (with class honours) and the M.Sc. degree from Federal University of Minas Gerais, Belo Horizonte, in 1988 and 1991, respectively. He received the Ph.D. degree in Control Systems Engineering from The University of Sheffield, Sheffield, U.K., in 1995. He holds a position of Full Professor, Department of Electronic Engineering, Federal University of Minas Gerais, Brazil. His research interests include system identification for nonlinear systems, NARMAX methods, model validation, prediction, spectral analysis, chaos, signal processing and neurosciences. \par

\vspace{1.2mm}

\begin{wrapfigure}{l}{25mm} 
\includegraphics[width=1in,height=1.25in,clip,keepaspectratio]{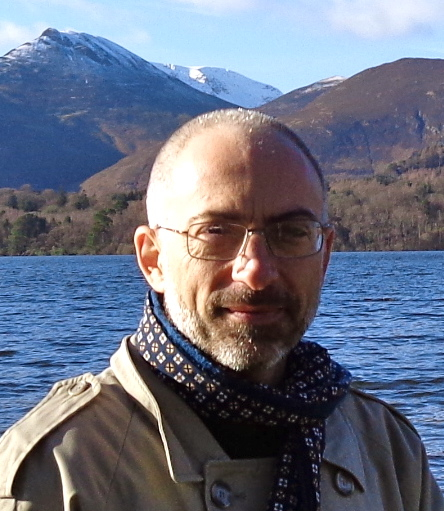}
\end{wrapfigure}\par
\textbf{Luis A. Aguirre} received a PhD degree in 1994 from the University of Sheffield, England. He joined the Department of Electronics Engineering at UFMG in 1995 where he currently serves as a full professor. He is the author of two books and was the Editor-in-Chief of Enciclopédia de Automática (3 volume set), sponsored by the Brazilian Society of Automation (SBA) and published by Editora Blücher. From 2009 to 2012 he served as the Editor-in-Chief of Controle \& Automação: Revista da Sociedade Brasileira de Automática, currently published by Springer Verlag under the name Journal of Control, Automation and Electrical Systems. His research includes the identification of nonlinear system, grey-box identification, nonlinear dynamics, and analysis of dynamical networks.
\par

\end{document}